\documentclass[pra,twocolumn,showpacs,preprintnumbers,amsmath,amssymb]{revtex4}
\usepackage{graphicx}% Include figure files
\usepackage{dcolumn}% Align table columns on decimal point
\usepackage{bm}% bold math
\usepackage{psfrag}
\usepackage{epsfig}
\usepackage{hyperref}
\expandafter\let\csname equation*\endcsname\relax
\expandafter\let\csname endequation*\endcsname\relax 
\usepackage{amsmath,bbm}
\usepackage{textcomp}
\usepackage{amssymb}
\usepackage{color}
\usepackage{mathbbol}
\usepackage[squaren]{SIunits}
\newcommand{\nn}{\nonumber}

\newcommand{\ehoch}{\ensuremath{\mathrm{e}^}}

\newcommand{\I}{\ensuremath{\mathrm{i}}}

\newcommand{\bra}[1]{\ensuremath{\langle#1|}}
\newcommand{\ket}[1]{\ensuremath{|#1\rangle}}

\newcommand{\expval}[1]{\ensuremath{\left\langle#1\right\rangle}}

\newcommand{\om}{\ensuremath{\omega_{\mathrm{m}}}}

\newcommand{\csound}{\ensuremath{c_{\mathrm{s}}}}
\newcommand{\gmj}{\ensuremath{g_{\mathrm{m},j}}}
\newcommand{\gmi}[1]{\ensuremath{g_{\mathrm{m},#1}}}

\newcommand{\gammam}{\ensuremath{\gamma_{\mathrm{m}}}}
\newcommand{\gammaeff}{\ensuremath{\gamma_{\mathrm{eff}}}}
\newcommand{\kappaex}{\ensuremath{\kappa_{\mathrm{ex}}}}
\newcommand{\xzpm}{\ensuremath{x_{\mathrm{ZPM}}}}
\newcommand{\xzpmi}[1]{\ensuremath{x_{\mathrm{ZPM},#1}}}

\newcommand{\bplus}{\ensuremath{b^{\dag}}}

\newcommand{\cnplus}{\ensuremath{c_n}^{\dag}}
\newcommand{\cn}{\ensuremath{c_n}}
\newcommand{\ac}{\ensuremath{a_{\mathrm{c}}}}

\renewcommand{\P}{\ensuremath{\mathcal{P}}}

\renewcommand{\L}{\ensuremath{\mathcal{L}}}
\renewcommand{\H}{\ensuremath{\mathcal{H}}}

\newcommand{\D}{\ensuremath{\mathcal{D}}}
\newcommand{\F}{\ensuremath{\mathcal{F}}}
\newcommand{\X}{\ensuremath{\mathcal{X}}}
\def\pM{\ensuremath{\stackrel{+}{\scriptstyle(\kern-1pt-\kern-1pt)}}}
\def\mP{\ensuremath{\genfrac{}{}{0pt}{1}{-}{\scriptstyle(\kern-1pt+\kern-1pt)}}}
\definecolor{mygray}{gray}{.30}

\newcommand{\sout}[1]{}  % Uncomment these two lines to switch off
\begin{document}

\title{Nonlinear Nanomechanical Resonators for Quantum Optoelectromechanics}
 \author{S.~Rips$^{1}$, I.~Wilson-Rae$^{2,1}$, and M.~J.~Hartmann$^{3,1}$}
\affiliation{$^{1}$Technische Universit{\"a}t M{\"u}nchen, Physik
  Department, James Franck Str., 85748 Garching, Germany,\\ 
$^{2}$Department of Physics, University of York, Heslington,
  York, YO10 5DD, United Kingdom,\\
$^{3}$Institute of Photonics and Quantum Sciences,
Heriot-Watt University, Edinburgh, EH14 4AS, United Kingdom.}

\date{\today}
\begin{abstract}
  We present a scheme for tuning and controlling nanomechanical
  resonators by subjecting them to electrostatic gradient fields,
  provided by nearby tip electrodes. We show that this approach
  enables access to a novel regime of optomechanics, where the
  intrinsic nonlinearity of the nanoresonator can be explored. In this
  regime, one or several laser driven cavity modes coupled to the
  nanoresonator and suitably adjusted gradient fields allow
    to control the motional state of the nanoresonator at the single
  phonon level. Some applications of this platform have been
  presented previously \cite{Rips2012,Rips2013}. Here, we provide a
  detailed description of the corresponding setup and its
    optomechanical coupling mechanisms, together with an in-depth
  analysis of possible sources of damping or decoherence and a
  discussion of the readout of the nanoresonator state.
\end{abstract}

%Classification Scheme.
\maketitle

%\tableofcontents
%\src{Table of contents rein oder raus?}
%

% ---------------------------------------------------------------------------
%
\section{Introduction}
\label{sec:intro}

Substantial progress in fabricating high-$Q$ mechanical resonators
with high frequencies, as well as recent success in cooling them close
to the motional ground state
\cite{nature08681,nature08967,nature10261,nature10461,Verhagen12},
inaugurate a new research field of manifold fundamental interest
\cite{science.1156032,Physics.2.40,PhysRevLett.88.148301}. The regime
of very low temperature, where a quantum mechanical description
predicts only few quanta of mechanical motion, promises potential
insight into some fundamental questions of decoherence, as well as
various technical applications that make use the of the expected
quantum behavior
\cite{nnano.2009.343,nphys1304,PhysRevA.82.061804}. Fundamental
questions, concerning the border between the classical (macroscopic)
and the quantum (microscopic) worlds \cite{Adler17072009}, trigger a
natural interest in preparing quantum states of ``as large as
possible'' objects and demonstrating their distinct quantum behavior
by appropriate measurements. See \cite{Aspelmeyer13} for a recent
review.

Regarding this major goal, it is important to stress that the dynamics
of a purely harmonic quantum system is analogous to its classical
dynamics, in the sense that expectation values of canonical
observables follow the classical equations of motion \cite{ehrenfest}.
Therefore, it is a common approach to introduce nonlinearities in a
quantum system in order to detect quantum behavior. While there may
be the possibility to achieve the strong optomechanical coupling regime
\cite{Rabl11,Nunnenkamp11,nphys1707} and make use of the nonlinear
nature of the standard optomechanical coupling, or to couple to a
nonlinear ancilla system
\cite{nature08967,PhysRevLett.88.148301,PhysRevLett.99.117203}, we
propose here a different approach: the use of an optoelectromechanical
system featuring a tunable \emph{mechanical} nonlinearity per
phonon. The latter originates from the intrinsic geometric
nonlinearity of elastic systems
\cite{Phys.Rev.B64_220101,1367-2630-8-2-021,1367-2630-10-10-105020,PhysRevLett.99.040404,PhysRevLett.101.200503,Dykman11}
and its amount per motional quanta is enhanced with the help of
electrostatic fields.  This has the advantage that the linear
optomechanical coupling is preserved as a control channel providing
techniques as, for example, the sideband driving technique used in
\cite{Rips2012}. The regime of large mechanical nonlinearity then
enables new means to control the mechanical motion at the quantum
level, if combined with the coupling to a high-Q optical cavity mode,
as well as the application of suitable gradient fields.

The intrinsic anharmonicity in the mechanical motion of micro-
  and nanomechanical resonators is usually small and therefore only
relevant in the regime of large oscillation amplitudes. In order to
render the anharmonicity relevant for displacements at the scale
of the quantum mechanical zero point motion, we propose to use
electrostatic gradient forces to enhance the latter
\cite{nature07932}. These forces result from the dielectric
  properties of the resonator material when an inhomogeneous external
  electric field is applied. They can be used to effectively
reduce the resonator's stiffness and therefore its resonance
frequencies. In turn, this has the effect that the zero point
deflection is enhanced up to an extent that the nonlinear
contribution becomes important.

Using this technique, the nonlinearity per phonon can be made large
enough, that distinct transitions in the mechanical spectrum can be
resonantly addressed while interacting with other quantum
systems. Examples are the selective sideband driving of transitions
%$\ket{n}\rightarrow\ket{n\pm1}$
in the mechanical spectrum \cite{Rips2012}, or the resonant exchange
of excitations within an array of nanoresonators via a common cavity
mode \cite{Rips2013}.

In this paper, we explicitly derive the fundamental mode properties of
a nonlinear mechanical resonator, subject to the aforementioned
gradient forces, to obtain a suitable model for the mechanical degree
of freedom. We then derive the specifics of the optomechanical
coupling to a high finesse cavity and analyze possible source of
damping and decoherence in detail. We also summarize different control
schemes, associated with suitable laser drives for the cavity and
gradient fields from the tip electrodes. Two applications of these
control mechanisms have been proposed previously
\cite{Rips2012,Rips2013}, considering state of the art experimental
components. We finally discuss a readout scheme for the nonlinear
  mechanical resonators considered.

The remainder of the paper is organized as follows. In section
\ref{sec:elasticity} we describe the nonlinear dynamics of thin rods
starting from elasticity theory and derive the resulting fundamental
mode Hamiltonian.  In section \ref{sec:softening} we describe our
approach to enhance the intrinsic mechanical nonlinearity by mode
softening with gradient forces. Then we introduce the optomechanical
model with several laser driven cavity modes in section
\ref{sec:optomech}. After introducing a possible setup for an
implementation in section \ref{sec:setup}, we quantitatively discuss
its central physical properties, namely the optomechanical coupling
mechanism in section \ref{sec:coupling}, as well as potential
setup specific losses in section \ref{sec:losses}. In section
\ref{sec:controlmech} we review some mechanisms that can be used to
control the mechanical motion at the level of single phonons and which
have been applied in previous works \cite{Rips2012,Rips2013}. Finally,
in section \ref{sec:measurement} we introduce methods to obtain
information on the mechanical state from the output spectrum of a
probe laser.

\section{Elasticity and fundamental mode description}
\label{sec:elasticity}

\begin{figure}
\includegraphics[width=\columnwidth]{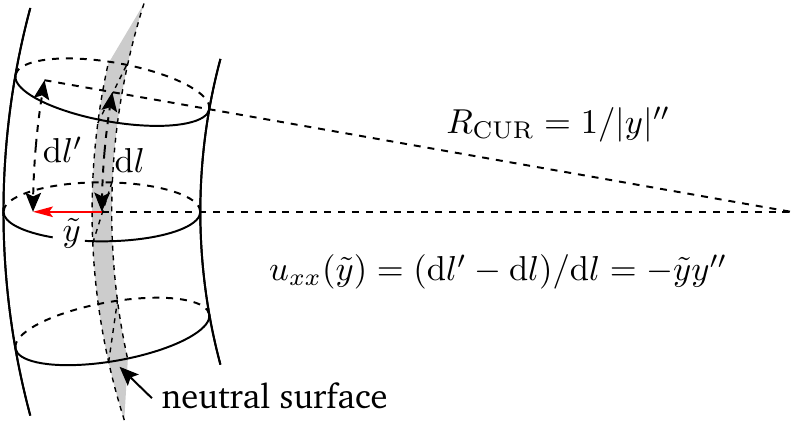}
\caption{Section of a deflected rod. The local strain $u_{xx}$
  depends on the transverse coordinate $\tilde y$ and on the
    local curvature $y''$ and determines the energy density
    $\tfrac{1}{2}Yu_{xx}^2$.}
\label{fig:bending_fig1}
\end{figure}

The harmonic description of the transverse motion of thin rods
is based on only considering the bending energy for small deflections
\cite{Landau-Lifshitz}. We consider here thin rods, which means that the
cross-sectional dimensions, such as the radius for circular cross
sections or width and depth for rectangular cross sections, are much
smaller than the length $L$. We also consider the rod to be
homogeneous along the longitudinal axis, here parametrized by
$x\in\{0,L\}$, with constant mass line density $\mu$ and use thin rod
elasticity theory. The planar deflection in the transverse direction
is described by a field $y(x)$ and we consider a bridge geometry
  where the end points at $x=0$ and $x=L$ are fixed,
  i.e.~$y(0)=y(L)=0$ and $y'(0)=y'(L)=0$. The Lagrangian within this
approximation reads
\begin{equation}
 \L(y(x,t))=\frac{\mu}{2}\int\mathrm{d}x\dot y^2-V_{\mathrm{b}}[y(x)]\,,
\label{Lagrangian}
\end{equation}
 with a kinetic part as well as the bending energy 
\begin{equation}
 V_{\mathrm{b}}[y(x)]=\frac{1}{2}\int\limits \mathcal{F}\tilde{\kappa}^2(y'')^2{\rm d}x\,.
\label{bending_energy}
\end{equation}
Here, $\mathcal{F}=Y A$ is the
linear modulus of the rod given by the Young modulus $Y$ of the
material times the cross-section area $A$, and
$\tilde{\kappa}^2=\frac{1}{A}\int_{\mathrm{cross}} \tilde y^2
{\rm d}A$ is the ratio between the bending and compressional
  rigidities and depends on the cross-sectional geometry, where
$\tilde y$ is the in-plane coordinate within the cross-section that is
directed along the deflection with origin at the neutral line, see
figure \ref{fig:bending_fig1}. For a rectangular cross-section of
thickness $d$, $\tilde{\kappa}=d/\sqrt{12}$, whereas for a
circular cross-section with radius $R$,
$\tilde{\kappa}=R/2$. For a cylindrical shell like a nanotube
one finds $\tilde{\kappa}=R/\sqrt{2}$. The energy
(\ref{bending_energy}) results from the fact that for small curvature
$y''$, the local strain inside the rod is linear with respect to the
distance $\tilde y$ to the neutral surface (cf.~Fig.~1) while the
free energy density is quadratic with respect to the strain.
The Lagrangian (\ref{Lagrangian}) leads to the equation of motion
\begin{equation}
% \mu \ddot{y}+\mathcal{F}\tilde{\kappa}^2y^{(4)}=0\,,
 \mu \partial_{t}^{2} y + \mathcal{F}\tilde{\kappa}^2 \partial_{x}^{4} y =0\,.
 \label{eqn_of_motion}
\end{equation}
%where $y^{(4)}$ denotes the $4$-th derivative with respect to $x$. 
As this equation is linear in $y$ and its derivatives, it leads to
harmonic dynamics characterized by the following eigenmodes
\begin{align}
  \phi_n(x)=\frac{1}{C_n}\bigg{[}&\frac{\sin(\nu_nx/L)-\sinh(\nu_nx/L)}{\sin(\nu_n)-\sinh(\nu_n)}\\ 
\label{modes}
&-\frac{\cos(\nu_nx/L)-\cosh(\nu_nx/L)}{\cos(\nu_n)-\cosh(\nu_n)}\bigg{]}\,,\nonumber
\end{align}
with frequencies $
\omega_n=\csound\tilde{\kappa}\left(\nu_n/L\right)^2$, where
$\csound=\sqrt{\F/\mu}$ is the phase velocity of compressional
phonons along the rod. The $\nu_n$ are the roots of the transcendental
equation $\cos\left(\nu_n\right)\cosh\left(\nu_n\right)\!=\!1$, with
$\nu_1=4.73$ the smallest one. The $C_n$ are normalization constants
chosen such that ${\mathrm{\bf
    max}}\left\{\phi_n(x)\right\}\!=\!1$. We choose this normalization
so that the coefficients in a mode expansion represent the
maximum amplitudes of the deflection associated to each
mode. Introducing now the canonical momentum $\Pi(x,t)=\delta
\L/\delta \dot y(x,t)$, as well as the expansion of the field into the
modes
\begin{equation}
 y(x,t)=\sum_n\phi_n(x)\mathcal{X}_n(t)\,,
\label{mode_expansion}
\end{equation}
yields the Hamilton function of a harmonic oscillator for each mode
\begin{equation}
\H_{\mathrm{lin}}=\sum_n\left(\frac{\mathcal{P}_n^2}{2m^*_n}+\frac{1}{2}m_n^*\omega_n^2\mathcal{X}_n^2\right)\,
\end{equation}
with the deflection $\X_n$ and mode momentum
$\P_n=m^*_n\partial_t\X_n$ for the $n$-th mode, as well as the
effective mode masses $m^*_n=\mu\int_0^L\phi^2_n(x){\rm d}x$.

Corrections to this harmonic description that lead to nonlinearities
originate from a stretching effect that occurs due to the deflection if
the end points of the rod are fixed
\cite{Phys.Rev.B64_220101}. The resulting strain leads to an
additional energy
\begin{equation}
  \frac{\mathcal{F}}{2}\frac{(L_t-L)^2}{L}\approx\frac{\mathcal{F}}{8L}\left(\int{\rm d}x(y')^2\right)^2\,,
\label{dynamic strain}
\end{equation}
where the stretched length is $L_t=\int\sqrt{1+(y')^2}{\rm d}x\approx L+\tfrac{1}{2}\int{\rm d}x(y')^2$ with $L$ being the zero deflection length. Including this streching energy leads to a nonlinear extension of the Hamiltonian, which after inserting the modes given in Eq.~\eqref{mode_expansion} leads to
\begin{equation}
 \H\!=\!\H_{\mathrm{lin}}+\frac{\mathcal{F}}{8L}\sum_{i,j,k,l}M_{ij}M_{kl}\mathcal{X}_i\mathcal{X}_j\mathcal{X}_k\mathcal{X}_l\,,
\label{H_multimode}
\end{equation}
where  
$
 M_{ij}=\int_0^L\phi_i'(x)\phi_j'(x){\rm d}x
$.
We quantize this model by introducing bosonic mode operators $\cnplus$ and $\cn$, given by
\begin{equation}
 c_n=\frac{1}{2x_{\mathrm{ZPM},n}}\, \mathcal{X}_{n} + \I \frac{x_{\mathrm{ZPM},n}}{\hbar}\, \mathcal{P}_{n}\,,
 \label{quantize_modes}
\end{equation}
where we introduced the zero point motion amplitudes $x_{\mathrm{ZPM},n}=\sqrt{\hbar/2m_n^*\omega_{n}}$ for each mode. This leads to the description
\begin{align}\label{Htot}
 \H=&\sum_n\hbar\omega_n\left(\cnplus\cn+\frac{1}{2}\right)\\
+&\hbar\sum\limits_{ijkl}\lambda_{ijkl}^0\left(c_i^{\dag}+c_i\right)\left(c_j^{\dag}+c_j\right)\left(c_k^{\dag}+c_k\right)\left(c_l^{\dag}+c_l\right),\nn
\end{align}
with nonlinearity
\begin{equation}
 \lambda_{ijkl}^0=\frac{\mathcal{F}}{8L\hbar}M_{ij}M_{kl}x_{\mathrm{ZPM},i}x_{\mathrm{ZPM},j}x_{\mathrm{ZPM},k}x_{\mathrm{ZPM},l}\,.
 \label{nonlinearity1}
\end{equation}
  As in a rigorous elasticity treatment this description arises from
  an adiabatic elimination of the stretching modes, the indices in
  Eq.~\eqref{Htot} should run up to an $N$ corresponding to an
  ``ultraviolet'' cutoff $\omega_N\sim\csound\pi/L$. The terms
  involving higher order modes induce small frequency shifts and
  nonlinear mode coupling. However, the later is found to be
  negligible for the parameters considered (see Appendix \ref{appendix
    B}) and the shift of the fundamental mode can be ignored given the
  additional tunable electrostatic contribution (see Section \ref{sec:softening}).

  Therefore we can restrict our description to the fundamental mode
  with $n=1$ and drop this label to get the usual Hamiltonian for the
  Duffing oscillator
\begin{equation}
 H_{\mathrm{m},0}=\frac{\P^2}{2m^*}+\frac{1}{2}m^*\omega_{\mathrm{m,0}}^2\mathcal{X}^2+\frac{\beta}{4}\X^4\,,
 \label{H_fundamental_mode}
\end{equation}
where we have introduced the fundamental frequency
$\omega_{\mathrm{m,0}}$ and the effective mass of the fundamental mode
$m^*\approx 0.3965\, \mu L$. The anharmonicity is given by
\begin{equation}
  \beta=\frac{\left(M_{11}L\right)^2}{2\nu_1^4\left(m^*/\mu L\right)}\,m^*\frac{\omega_{\mathrm{m},0}^2}{\tilde{\kappa}^{2}}\approx0.060\,m^*\frac{\omega_{\mathrm{m},0}^2}{\tilde{\kappa}^{2}}\,.
\end{equation}
In terms of phonon creation and annihilation operators $b^{\dag} = c_{1}^{\dag}$ and $b = c_{1}$ this Hamiltonian reads,
\begin{equation}
  H_{\mathrm{m},0}=\hbar\omega_{\mathrm{m},0}\bplus b +\hbar\frac{\lambda_0}{2}\left(\bplus+b\right)^4\,,
 \label{Hm0}
\end{equation}
with the nonlinearity parameter
$\lambda_0\equiv2\lambda_{1111}^0=\frac{\beta}{2}\xzpm^4/\hbar$.
Here, the frequencies $\omega_{\mathrm{m},0}$ and $\lambda_0$ refer to
fundamental mode properties that result only from the intrinsic
elastic forces in the absence of any externally applied forces on the
rod. As we will describe in the next section, external forces can be
used to tune the resonance frequency,
$\omega_{\mathrm{m},0}\rightarrow \om$, which will in turn change the
zero point motion amplitude, $\xzpm$, and hence the nonlinearity,
$\lambda_0\rightarrow \lambda$.

%\section{Enhanced nonlinearity by mode softening}
\section{Electric gradient fields}

\label{sec:softening}

\begin{figure}
\centering
\includegraphics[width=0.9\columnwidth]{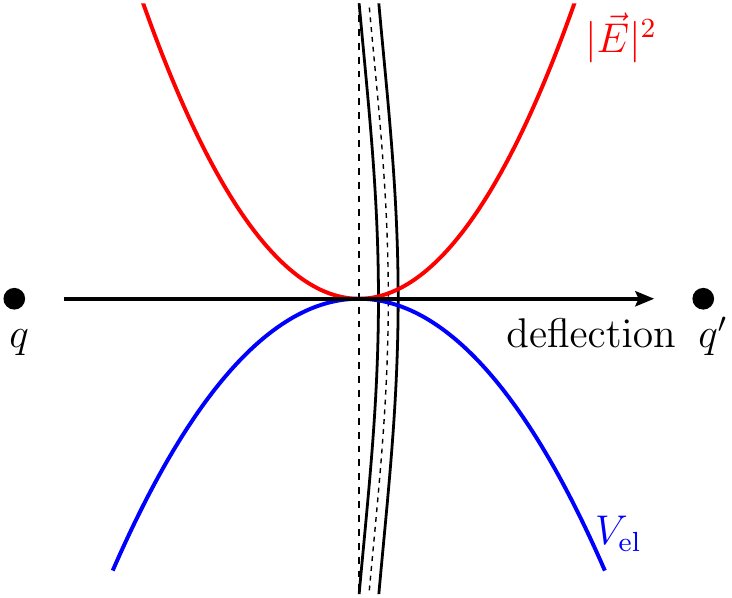}
\caption{Resonator with electrodes modeled by point charges $q,q'$. The field profile (red) leads to an inverted parabola for the dielectric potential (blue) around the equilibrium position. }
\label{fig:bending_fig2}
\end{figure}

In this section we describe how electric fields generated by tip
electrodes that are placed near the center of the doubly clamped
nanobeam, see figure \ref{fig:bending_fig2}, can be employed to
control its dynamical properties. In particular, inhomogeneous
  gradient fields can be used to enhance the nonlinearity per
  phonon. The later scales as
$\lambda\propto\xzpm^4\propto\omega_{\mathrm{m}}^{-2}$ and can thus be
enhanced by lowering the harmonic oscillation frequency
$\omega_{\mathrm{m},0}$. One way to change the mode frequency that has
been discussed previously, is to add an additional external force
along the rod axis, that causes compressive or tensile strain
\cite{Phys.Rev.B64_220101, Europhys.Lett.65_158-164}. An alternative
approach, that promises better control but yields the same potential
for the fundamental mode, is to use a static electric field that is
strongly inhomogeneous in the direction of deflection. If the rod
material shows suitable dielectric properties, this produces an
additional, inverted square potential with respect to the deflection,
see figure \ref{fig:bending_fig2}.

Here we consider tip electrodes with suitable applied voltages at each
side of the nanoresonator that generate an electrostatic field.  In
our calculation we model the electrodes by point charges $q$ and
  $q'$, which is valid given that the relevant tip radii are much
  smaller than the gap between the electrodes (cf.~Section \ref{sec:setup}). The
electrostatic energy associated to the dielectric per unit length
along the rod can be described by
 \begin{equation}
 W(x,y)=-\frac{1}{2}[\alpha_{\parallel}E^2_{\parallel}(x,y)+\alpha_{\perp}E^2_{\perp}(x,y)]\,,
\end{equation}
where $x,y$ are the co-ordinates along the resonator axis and the direction of its deflection. $E_{\parallel,\perp}$ are external field components parallel and perpendicular to the beam axis and 
$\alpha_{\parallel,\perp}$ the respective screened polarizabilities. We can expand $W(x,y)$ to second order in the displacement $y$ and get an additional contribution to the Hamiltonian of the nanobeam that reads
\begin{align}
V_{\mathrm{el}}=&V_{\mathrm{el}}^{(1)}+V_{\mathrm{el}}^{(2)}\\
=&\int\limits_0^L \left[\left.\frac{\partial W}{\partial y}\right\vert_{y=0}y+\frac{1}{2}\left.\frac{\partial^2 W}{\partial y^2}\right\vert_{y=0}y^2\right]{\rm d}x\,,\nonumber
\end{align}
where we dropped the displacement independent constant $W(x,0)$ which is irrelevant for the dynamics. Inserting the modes defined in equation \eqref{mode_expansion} we get 
\begin{equation}
 V_{\mathrm{el}}=\sum_nF_n\mathcal{X}_n+\frac{1}{2}\sum_{lk}W_{lk}\mathcal{X}_l\mathcal{X}_k\,,
\end{equation}
with 
\begin{align}
 F_n&=\int\limits_0^L\left.\partial_yW(x,y)\right\vert_{y=0}\phi_n{\rm d}x\,,\\W_{lk}&=\int\limits_0^L\left.\partial_y^2 W(x,y)\right\vert_{y=0}\phi_l\phi_k{\rm d}x\,.
\end{align}
The contributions $W_{lk}$ will cause weak interactions between the
modes, as $W_{lk}$ is not a diagonal matrix. By diagonalizing the
Hamiltonian of the rod in the presence of electrostatic fields, one
can find new normal modes. For the parameters considered in this work,
the induced corrections to the mode shapes are however found to be
negligibly small. We thus focus on the fundamental mode contributions
$F_{0}$ and $W_{00}$, where $W_{00}<0$.

One may also consider time dependent electric fields and it is convenient to separate between static and time dependent contributions, 
\begin{align}
 F_0&=F_0^{\mathrm{s}}+F_0(t)\label{Fn}\\
 W_{00}&=W_{00}^{\mathrm{s}}+W_{00}(t)\,. 
\end{align}
Whereas the time dependent contributions $F_{0}(t)$ and $W_{00}(t)$
generate drives applied to the nanobeam and will be considered in
section \ref{sec:controlmech}, we now focus on the constant
contributions $F_0^{\mathrm{s}}$ and $W_{00}^{\mathrm{s}}$, which can
be employed to tune the nanobeam.

The electrostatic force $F_0^{\mathrm{s}}$ causes a static deflection
of the nanobeam and shifts its equilibrium position. However if the
rod interacts with the photon fields of a nearby cavity, those fields
will also cause a deflecting force. For convenience we choose
$F_0^{\mathrm{s}}$ such that these two forces compensate and the
equilibrium position remains unshifted (see section
\ref{sec:optomech}).  The electrostatic potential associated to
$W_{00}^{\mathrm{s}}$ in turn is an inverted harmonic potential that
lowers the harmonic oscillation frequency of the nanobeam. Therefore,
we consider
\begin{equation}
 H_{\mathrm{m}}=\frac{\mathcal{P}^2}{2m^*}+\frac{1}{2}m^*\om^2\mathcal{X}^2+\frac{\beta}{4}\mathcal{X}^4,
\end{equation}
as the ``tuned'' mechanical Hamiltonian with a reduced frequency
$\om^2\approx\omega_{\mathrm{m},0}^2-|W_{00}^{\mathrm{s}}|/m^*$. 
In a phononic description this Hamiltonian reads
\begin{equation}
 H_{\mathrm{m}}=\hbar\om\bplus
 b+\hbar\frac{\lambda}{2}(\bplus+b)^4\,,%+F_0^{\mathrm{s}}\xzpm(\bplus+b)\,, 
 \label{Hm1}
\end{equation}
where the nonlinearity per phonon $\lambda=\zeta^2\lambda_0$ is now
increased by a factor
$\zeta^2=\left(\omega_{\mathrm{m},0}/\om\right)^2$,
cf.~Eq.~\eqref{Hm0}. As an example, for maximum applied fields at the
tube $E_\parallel\approx1.2\times10^{7}\,$Vm$^{-1}$ and
$E_\perp\approx1.8\times10^{6}\,$Vm$^{-1}$ and a gap between the
electrodes of size $D=40\,$nm (cf.~Fig.~3), typical parameters
discussed in Section \ref{sec:setup}, yield $\zeta>3$ allowing to
boost $\lambda$ by at least an order of magnitude.

For further calculations, it is convenient to express all observables
in the energy eigenbasis of the Hamiltonian $H_{\mathrm{m}}$, so that
\begin{equation} \label{eq:diagHm}
 H_{\mathrm{m}}=\sum\limits_nE_n\ket{n}\bra{n}\,,\qquad
 \mathcal{X}/\xzpm=\sum\limits_{nm}X_{nm}\ket{n}\bra{m}\,, 
\end{equation}
where the energy eigenstates $\left\{\ket{n}\right\}$ and energy
levels $E_n$, as well as the displacement matrix elements $X_{nm}$
have to be determined numerically.  For small nonlinearites
$\lambda\ll\om$ analytical expression can be obtained as the
Hamiltonian \eqref{Hm1} is approximately diagonal in Fock basis since
one may apply a rotating wave approximation in the nonlinear part
\begin{equation}
 H_{\mathrm{m}}\rightarrow H_{\mathrm{m}}'=\hbar\om'\bplus
 b+\hbar\frac{\lambda'}{2}\bplus\bplus bb\,, 
\end{equation}
where $\om'= \om+2\lambda'$, $\lambda'=6\lambda$ and the
eigen-energies are given by $E_n\approx
n\om'+\frac{n(n-1)}{2}\lambda'$.

The time dependent contributions $F_0(t)\X$ and $W_{00}(t)\X^2$ in
turn can be used to drive or temporarily detune the resonator. This
has been used in \cite{Rips2013} to perform local gate operations on
nanobeams acting as qubits.

In the following section we will now discuss an optomechanical
interaction between the nano resonator described by the Hamiltonian in
equation (\ref{Hm1}) and the resonance modes of a high finesse cavity.

\section{Optomechanical Model}
\label{sec:optomech}

We consider a typical optomechanical setup with a micro toroid cavity
coupled to several nanomechanical resonators, where the displacement
of the latter modifies the frequencies of the cavity modes. The cavity
modes with frequencies $\omega_i$ that contribute to the dynamics are
described by photon creation and annihilation operators ${a_i}^{\dag}$
and ${a_i}$. They are each driven by a classical laser of input power
$P_i$ and frequency $\omega_{\mathrm{L},i}$. The coupling strength
between cavity mode $i$ and nanobeam $j$ is given by
$G_{0,ij}\xzpmi{j}$, where $G_{0,ij}=\tfrac{\partial
  \omega_i}{\partial\X_j}$ is the optical frequency shift per
deflection $\X_j$. Thus, in a frame rotating with the laser field
modes, the Hamiltonian describing the coupled system reads
($\hbar\rightarrow1$)
\begin{align}
H=&\sum\limits_i\left[-\Delta_i{a_i}^{\dag}a_i+\frac{\Omega_i}{2}\left(a_i^{\dag}+a_i\right)\right]+\sum\limits_jH_{\mathrm{m},j}\nonumber\\
&+\sum\limits_{ij}G_{0,ij}{a_i}^{\dag}a_i\X_j\,.
\label{hamiltonian}
\end{align}
Here we have introduced the laser detunings denoted by $\Delta_i=\omega_{\mathrm{L},i}-\omega_i$ and drive amplitudes $\Omega_i/2=\sqrt{P_i\kappa_{\mathrm{ex},i}/\hbar\omega_{\mathrm{L},i}}$, with $\kappa_{\mathrm{ex},i}$ being the cavity decay rates into the associated outgoing electromagnetic modes. 

Both the field inside the cavity and the mechanical motion are subject
to damping, which in the regime of weak optomechanical coupling, small
nonlinearity and low mechanical occupation is well described by a
master equation with Lindblad form damping terms. With the decay rates
for the cavity modes $\kappa_i$ and the mechanical damping rates
$\gamma_j$, this master equation reads
\begin{align}
\label{mg}
 \dot\rho=-\I\left[H,\rho\right]&+\sum\limits_i\frac{\kappa_i}{2}\D(a_i)\rho\,+\\
 \nonumber&+\sum\limits_j\frac{\gamma_j}{2}\left[(\overline n_j+1)\D(b_j)\rho+\overline n_j \D(\bplus_j)\rho\right]\,.
 \end{align}
 Here we introduced the Lindblad form dissipator $\D(\hat o
 )\rho=2\hat o \rho \hat o^{\dag} -\hat o^{\dag} \hat o \rho-\rho\hat
 o^{\dag}\hat o$, as well as the Bose occupation number $\overline
 n_j=\left[\exp\left(\frac{\hbar\omega_{\text{m},j}}{k_{\mathrm{B}}T}\right)-1\right]^{-1}$
 of the phonon bath mode with frequency $\omega_{\text{m},j}$ at
 temperature $T$. A more precise treatment of the mechanical damping
 accounts for the fact that, due to the mechanical nonlinearity, there
 is more than one bath mode resonantly coupling to the mechanical
 mode. However, for small nonlinearity, equation (\ref{mg})
 proofs to be sufficiently accurate.
 
 As usual, we expand the cavity field operators around their steady
 state values and adopt a shifted representation $a_i\rightarrow
 a_i+\alpha_i$, with $\alpha_i=\Omega_i/(2\Delta_i+\I\kappa_i)$, in
 which the master equation has the same form as in equation (\ref{mg})
 but with the shifted system Hamiltonian
\begin{align}
\nonumber H'=&-\sum\limits_i\Delta_i{a_i}^{\dag}a_i+\sum\limits_jH_{\mathrm{m},j}\\
 &+\sum\limits_{ij}\left(\frac{\gmi{ij}^*}{2}a_i+\text{H.c.}\right)\left(\bplus_j+b_j\right)
 \label{hamiltonian_shifted}
\end{align}
with $\gmi{ij}=2\alpha_iG_{0,ij}\xzpmi{j}$ and where we have
dropped the nonlinear terms $\propto {a_i}^ {\dag}a_i(\bplus_j+b_j)$
in the coupling, which is valid for $\expval{{a_i}^
  {\dag}a_i}\ll|\alpha_i|^ 2$. We have also assumed that the
static electric fields for each beam are chosen such that
$F_{0,j}^{\mathrm{s}}=-\hbar\sum_iG_{0,i}|\alpha_i|^2$, so that their
equilibrium positions are undeflected.

We now turn to discuss a possible experimental setup that would allow
to explore the physics described by the model presented in equation
\eqref{mg}.

\section{Setup}
\label{sec:setup}
\begin{figure}
\centering
\includegraphics[width=\columnwidth]{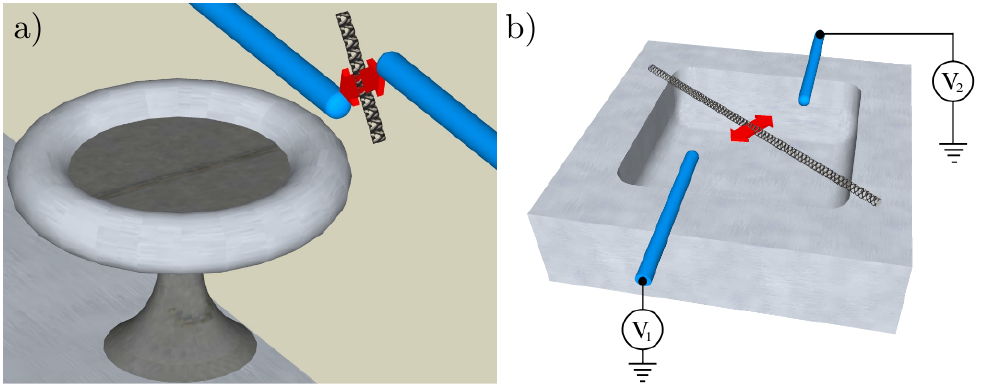} \\
\raisebox{0.5cm}{\includegraphics[width=0.4\columnwidth]{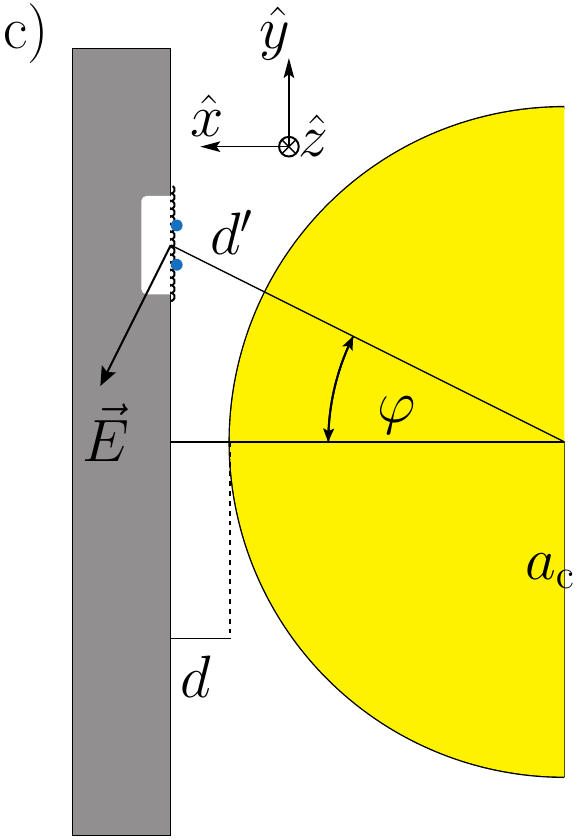}} \hspace{.5cm}
\raisebox{0.55cm}{\includegraphics[width=0.4\columnwidth]{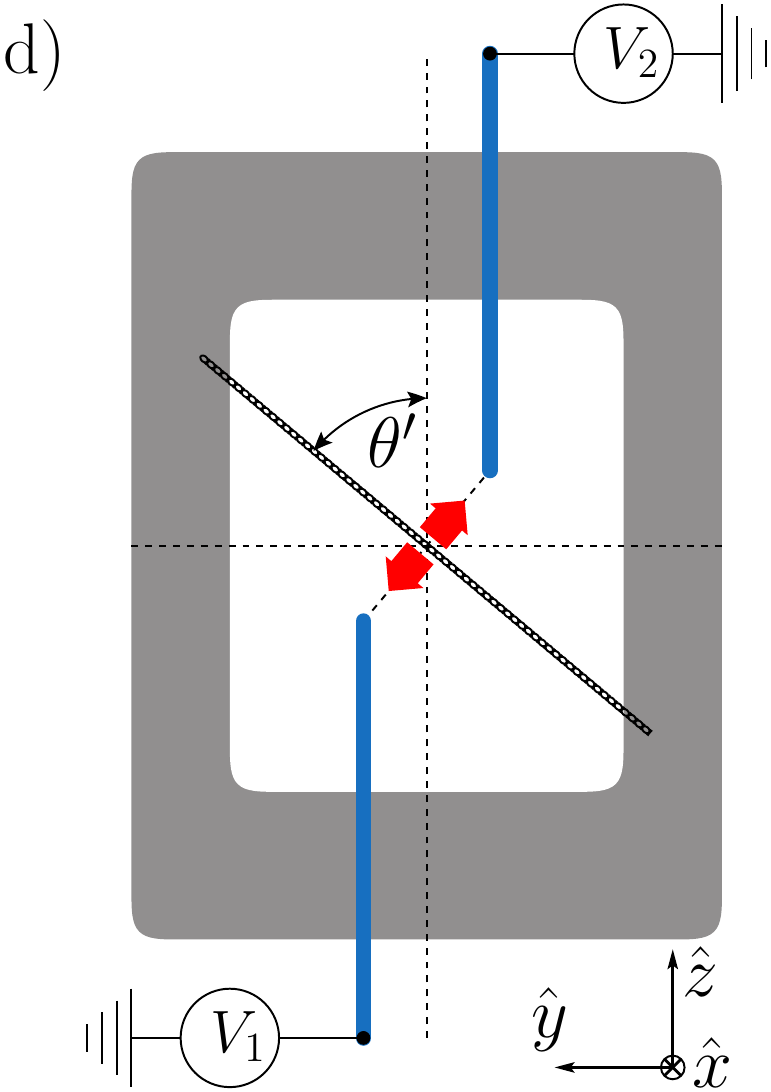}}
\caption{a) Microtoroid cavity with nearby CNT (not to scale). b)
  NEMS-chip with CNT-resonator and electrodes. c) Schematic top view
  of the NEMS chips with the CNT oscillating in the $\hat z$-$\hat
  y$-plane. d) Schematic cross-sectional view of the NEMS-Chip and
  the cavity rim (yellow). The NEMS-chip is mounted in the evanescent
  cavity field at a distance $d$ to the cavity rim surface. The
  nanoresonator is displaced from the closest point to allow for an
  optomechanical coupling that is linear in the resonator's
  deflection.}
\label{fig:setup}
\end{figure}

For the experimental realization of the model in equation
\eqref{mg}, we envisage a setup as shown in Fig.~\ref{fig:setup},
  comprising a NEMS chip containing the nanobeam resonators and a
high finesse toroidal microcavity
\cite{nphys1425,PhysRevLett.103.053901,Lee10}.
Each nanobeam resonator consists of a suspended single-walled
  carbon nanotube (CNT) with radius $R$
  ($c_\mathrm{s}=2.1\times10^{4}\,$ms$^{-1}$, $\mu=(2\pi
  R)\times7.6\times10^{-7}\,$kgm$^{-2}$)
  \cite{science.1176076,science.1174290}, and has electrodes in its
  vicinity that generate the electric fields for controlling and
  driving it. Furthermore, the nanoresonator interacts with the
  evanescent field of the microcavity via optical gradient forces
(see section \ref{sec:coupling}). Given the state of the art, a
  CNT is a favorable system to implement the proposed
  optoelectromechanical scheme \cite{Hofmann13,Waissman13,Stapfner13}. In
  particular, the intrinsic nonlinearity per phonon scales as
  $\lambda_0\propto\left(\tilde{\kappa}^2m\right)^{-1}$, which
  favors small transverse dimensions and masses. Additionally, carbon
  nanotubes show ultra-low dissipation \cite{doi:10.1021/nl900612h},
  that is expected to decrease further in the regime of small
  amplitudes \cite{nnano.2011.71}. It is convenient to use ($10$,$0$)
  CNTs with radius $R=\unit{0.39}{\nano\meter}$, as these tubes show
relatively large static polarizabilities with
$\alpha_{||}=143\left(4\pi\epsilon_0\text{\AA{}}{}^2\right)$,
$\alpha_{\perp}=10.9\left(4\pi\epsilon_0\text{\AA{}}{}^2\right)$
\cite{PhysRevLett.96.166801}, where $\alpha_{||}$ ($\alpha_{\perp}$)
is the polarizability parallel (perpendicular) to the tube
axis. To obtain nonlinearities that are large enough, it is
  convenient to use CNT lengths below $\unit{1}{\micro\meter}$ ---for
  $L=\unit{1}{\micro\meter}$ we have
  $\omega_{\mathrm{m},0}=\unit{20.6}{\mega\hertz}$ and
  $\lambda_0=\unit{2.24}{\kilo\hertz}$.

  The cavity is a silica ($n_{\mathrm{c}}=1.44$) microtoroid with
  resonant wavelengths $2\pi
  c/\omega_j\approx\lambda_c=\unit{1.1}{\micro\meter}$, circumference
  $L_{\mathrm{c}}\sim\unit{1}{\milli\meter}$, finesse
  $\mathcal{F}_{\mathrm{c}}=3\times10^6$ \cite{kippenberg:6113} and
  mode volume $V_{\mathrm{c}}\approx L_{\mathrm{c}}\times
  \unit{6}{\square\micro\meter}$. These parameters correspond to
  $a_c\approx \unit{2.0}{\micro\meter}$, $\xi\approx0.2$ and a decay
  length of the evanescent field
  $\kappa_{\perp}^{-1}\approx\unit{0.17}{\micro\meter}$
  [cf.~Eq.~\eqref{xi} in Section \ref{sec:coupling}]. The NEMS chip is
  placed at a distance $d=\unit{50}{\nano\meter}$ from the cavity
  rim. The CNT-resonator is displaced from the closest point to allow
  for a linear coupling, as the resonator moves in the plane of the
  chip's surface (cf.~Section \ref{sec:coupling}), and the electrodes
  are aligned parallel to the rim of the cavity to minimize additional
  cavity losses they might induce, see section \ref{sec:losses}.

In the following two sections we analyze important practical aspects
of the envisioned implementation with a carbon nanotube coupled to a
toroidal microcavity in more detail. Thus, readers who are only interested
in the results of the discussed mechanism may directly turn to
section \ref{sec:controlmech}.

\section{Optomechanical coupling}
\label{sec:coupling}

In this section we derive an estimate for the coupling between the
mechanical displacement and the cavity field. The coupling arises as
the energy of the dielectric oscillator in the evanescent electric
field depends on the field strength at the location of the
oscillator. As the evanescent field decays with distance to the cavity
rim, altering the oscillator-cavity distance by displacing the
oscillator results in a change of energy. Thus, the coupling part of
the Hamiltonian is given by
\begin{equation} \label{eq:couplingham}
H_{\mathrm{c}}=-\frac{1}{2}\int_{V_R}\vec P(\vec r)\cdot \vec E(\vec r) {\rm d}V
\end{equation}
where the polarization is $\vec P(\vec
r)=\overleftrightarrow{\alpha}\cdot\vec E(\vec r)$ with
$\overleftrightarrow{\bf\alpha}$ the screened polarizability tensor
and $\vec E$ the cavity field. The integration is taken over the
oscillator volume $V_R$ and without loss of generality we
consider a single cavity mode with resonant frequency $\omega_{\mathrm{c}}$.

Given the dimensions of the toroidal microcavity, its torus can be
locally modeled as a cylindrical waveguide of radius $\ac$ ---note
that this differs from the definition of $\ac$ used in
Ref.~\onlinecite{Rips2012} by a factor of $1.44$ [cf.~below
Eq.~\eqref{U_r_2}]. We introduce cylindrical coordinates $(r,\phi,z)$
with the $\hat z$-direction along the waveguide axis
(cf.~Fig.~\ref{fig:setup}c) and consider TE$_{0,1}$ modes, as a
transverse electric field is advantageous to suppress loss mechanisms
that are discussed in section \ref{sec:losses}. The corresponding
transverse fields are given by \cite{Jackson8}
\begin{align}
B_{r}=&\I\frac{k_{||}}{\gamma^2}\frac{\partial B_{z}}{\partial r}\,,\nonumber\\
 E_{\varphi}=&-\frac{\omega_{\mathrm{c}}}{k_{||}}B_{r}\,, \label{inside}
\end{align}
for the field inside the waveguide, $r<\ac$, and 
\begin{align}
B_{r}=&-\I\frac{k_{||}}{\kappa_{\bot}^2}\frac{\partial B_{z}}{\partial r}\,,\\
E_{\varphi}=&-\frac{\omega_{\mathrm{c}}}{k_{||}}B_{r}\,,
\end{align}
outside the waveguide, $r>\ac$. The axial field reads
\begin{align}
B_{z}(r,z)&=B_{z}(0)J_0(\gamma r)\ehoch{\I k_{||}z}\,, &r<\ac\,,\label{axial}\\
B_{z}(r,z)&=\tilde\xi
B_{z}(0)\frac{K_0^{(1)}(\kappa_{\bot}r)}{K_0^{(1)}(\kappa_{\bot}\ac)}\ehoch{\I
  k_{||}z}\,, &r>\ac\,, 
\end{align}
with the modified Bessel function $K_{0}^{(1)}$ and the Bessel
function of the first kind $J_{0}$. Here
$\tilde\xi=B_{z}(\ac,z)/B_{z}(0,z)$, $k_{||}$ is the wavevector
component parallel to the waveguide axis and $\I\kappa_{\bot}$ and
$\gamma$ are the transverse wave vectors outside and inside the
waveguide, respectively. Henceforth, we assume a refractive index such
that $n_{\mathrm{c}}^2-1\sim n_{\mathrm{c}}^2$ and a frequency
$\omega_{\mathrm{c}}$ well above cutoff,
i.e.~$(n_{\mathrm{c}}k\ac/x_{1,1})^2\gg1$ ---where
$k=\omega_{\mathrm{c}}/c$ and $x_{1,1}\approx3.8$ is the first zero of
$J_1(x)$. These assumptions imply that $k_{||}\approx
n_{\mathrm{c}}k$, $\kappa_\perp\approx \sqrt{n_{\mathrm{c}}^2-1}k$,
$\gamma\ll\kappa_\perp$, $\kappa_{\perp}\ac\gg1$, and
$\ac\gamma\approx x_{1,1}$. Within these approximations, the ratio of
the axial magnetic field at $r=\ac$ to its value at the origin is
given by $\tilde\xi=J_0(x_{1,1})\approx-0.4$, and the evanescent field
can be written as
\begin{equation}
 E_{\varphi}(r)\approx-\I\frac{\omega_{\mathrm{c}}}{\kappa_{\bot}}\tilde\xi
 B_{z}(0)\sqrt{\frac{\ac}{r}}\ehoch{-\kappa_{\bot}(r-\ac)}\ehoch{\I
   nkz}\,. 
\label{E_phi}
\end{equation}
This field will later be used to estimate losses induced by the
electrodes. In order to determine the optomechanical coupling
strength, we write the electric field in its quantized form
\begin{equation}\label{Equantized}
  \vec E(\vec r) = \sqrt{\frac{\hbar
      \omega_{\mathrm{c}}}{2\epsilon_0}}(a^{\dag}+a) u_\varphi(\vec
    r)\hat\varphi\,, 
 \end{equation}
with photon creation (annihilation)
 operators $a^{\dag} (a)$ and where $\vec u(\vec r)\propto\vec
   E(\vec r)$ is the corresponding normalized eigenmode, satisfying
\begin{equation}\label{eigenmode}
  \int \frac{\epsilon(\vec r)}{\epsilon_0}\left|\vec u(\vec
    r)\right|^2{\rm d}V =1\,.
\end{equation}
We consider the following definition for the mode volume
\begin{equation}\label{mode-volume}
  V_{\mathrm{c}}=\int{\rm d}V\frac{\epsilon(\vec r)|\vec E(\vec
    r)|^2}{n_{\mathrm{c}}^2\epsilon_0|\vec E_{\mathrm{max}}|^2}\,, 
\end{equation}
From Eqs.~\eqref{inside} and \eqref{axial} and properties of the
  Bessel functions, we obtain that the maximum electric field inside
  the cavity is given by
\begin{equation}\label{Emax}
 |\vec E_{\mathrm{max}}|=\frac{\omega_{\mathrm{c}}}{\gamma}\left|B_z(0)\right|
J_1\left(x_*\right)\approx\frac{\omega_{\mathrm{c}}}{\gamma}
\frac{\left|B_z(0)\right|}{1.7}\,, 
\end{equation}
where $x_*$ is the first positive root of $J_0(x)=J_2(x)$. Thus,
neglecting the small contributions of the evanescent part to the
integrations in Eqs.~\eqref{eigenmode} and \eqref{mode-volume} which
are higher order in $1/\kappa_\perp\ac$, we obtain from
Eqs.~\eqref{E_phi}-\eqref{Emax}
\begin{equation}
  u_\varphi(\vec r)\approx
  \frac{-\xi}{n_{\mathrm{c}}\sqrt{V_{\mathrm{c}}}}\sqrt{\frac{\ac}{r}}
  \,\ehoch{-\kappa_{\bot}(r-\ac)}\, 
 \label{U_r_2}
\end{equation}
for $r>\ac$. Here, $V_{\mathrm{c}}\approx 0.5\,\pi\ac^2L_{\mathrm{c}}$ and
\begin{equation}\label{xi}
  \xi=\frac{\gamma|\tilde\xi|}{\kappa_{\perp}J_1(x_*)}=
  \frac{\lambda_{\mathrm{c}} x_{1,1}\left|J_0(x_{1,1})\right|}{2\pi
    \ac\sqrt{n_{\mathrm{c}}^2-1}J_1(x_*)}\approx \frac{0.42\,
    \lambda_{\mathrm{c}}}{\ac\sqrt{n_{\mathrm{c}}^2-1}}
\end{equation}
denotes the ratio of the electric field at the waveguide's surface to
the maximum field $E_{\mathrm{max}}$.

The zero point motion $\xzpm$ of the oscillator is small compared to
the decay length $1/\kappa_{\bot}$ of the evanescent field and
the same holds for the transverse dimensions of the
oscillator. Thus, we can linearize $\vec E(\vec
  r)\cdot\overleftrightarrow{\alpha}\cdot\vec E(\vec r)$ around the
  equilibrium position of the nanoresonator and assume that the
  electric field is constant everywhere inside the resonator volume
  $V_R$. Subsequently, by comparing the Hamiltonians
  \eqref{eq:couplingham} and \eqref{hamiltonian}, and using
  Eqs.~\eqref{Equantized} and \eqref{U_r_2}, we find for the
  opto-mechanical coupling rate,
\begin{equation} \label{eq:couplingexpress}
G_0\approx\frac{\omega_{\mathrm{c}}\alpha_{||}\kappa_{\bot}L\xi^2}{n_{\mathrm{c}}^2
\epsilon_0V_{\mathrm{c}}}\ehoch{-2\kappa_{\bot}d} C_{\mathrm{corr}}\,.
\end{equation}
Here, we have neglected the contribution of the perpendicular
polarizability since, for carbon nanotubes, the perpendicular
polarizability is typically one order of magnitude smaller than the
parallel one, and again used $\kappa_\perp\ac\gg1$ so that only
  the derivative of the exponential factor is relevant. The geometry
of the setup we consider here is illustrated in figure \ref{fig:setup}
and the dependence on the alignment and positioning of the
nanotube is accounted for in the correction factor
$C_{\mathrm{corr}}$.  For a TE$_{0,1}$ mode of the cavity field,
the electric field is directed along $\hat\varphi$, i.e. tangential to
the cavity rim, while the nanotube is aligned along $\vec
e_{\mathrm{nt}}=\cos\theta'\hat z+\sin\theta'\hat y$.  In addition,
the deflection of the nanotube in the direction $\vec
e_{\mathrm{osc}}=\cos\theta'\hat y-\sin\theta'\hat z$ is not aligned
with the interaction-energy gradient, which is approximately
along $\hat r=\cos\varphi\hat x+\sin\varphi\hat y$. Finally, if $d$ is
the distance of the chip to the cavity rim, the actual distance of the
oscillator to the rim is $d'=(\ac+d)/\cos\varphi-\ac$. Taking
into account these various issues, we find for the correction
factor
\begin{equation}
C_{\mathrm{corr}}\approx\ehoch{-2\kappa_{\bot}(d+\ac)(\text{sec}\varphi-1)}
\sin^2\theta'\cos\theta'\cos^2\varphi\sin\varphi\,.
\end{equation}
This is maximized for $\sin^2\theta'_*=2/3$ and
$\varphi_*\approx1/\sqrt{2\kappa_{\bot}(d+\ac)}$, where we consider
the leading order in the small parameter
$1/2\kappa_\perp(d+\ac)$. These optimal angles yield
$C_{\mathrm{corr}}\approx 0.17/\sqrt{\kappa_{\bot}(d+\ac)}$, resulting
in $1/C_{\mathrm{corr}}\approx22$ for the parameters introduced in
Section \ref{sec:setup} (i.e.~$n_{\mathrm{c}}=1.44$ and
$d\ll\ac\approx\unit{2.0}{\micro\meter}$). Finally, for those typical
values we obtain $G_0\approx 1.02 \times 10^{10} \hertz / \meter$.

\section{Loss Mechanisms}
\label{sec:losses}

In addition to the well known loss mechanisms of photon losses from
the cavity and intrinsic phonon losses of the mechanical resonator
\cite{Wilson-Rae08,Remus09}, there can be further sources of loss in our
setup due to the presence of the tip electrodes.  In this section we
show that these additional loss mechanisms are negligible for the
parameters we envision.

\subsection{Cavity losses induced by metallic nanotube electrodes}
\label{electrode losses}

Exploiting the nonlinearity of the nanoresonators to control their
dynamics demands a high-$Q$ optical cavity as the linewidth $\kappa$
needs to be at most comparable to the mechanical nonlinearity
$\lambda$. Using conventional metallic electrodes to generate the
inhomogeneous control fields can potentially increase the cavity
losses. To achieve the necessary low losses it is crucial to have deep
subwavelength transverse dimensions for the electrodes.

We now give an estimate of the photon losses that may be induced by
such electrodes and show that these are negligible.  To do so, we
model the electrodes as metallic cylinders and assume that their
radius $R'$ is much smaller than the decay length of the
evanescent cavity field, $R'\ll\kappa_{\bot}^ {-1}$. We assume
the electrodes to be parallel to the waveguide representing the cavity
rim, with a small misalignment angle $\theta$. For the relevant
  TE$_{0,n}$ modes, the losses result solely from the misalignment, as
  in the small radius regime considered they arise only from the field
  along the electrode axis which vanishes for $\theta=0$. The
resulting finesse $\mathcal{F}$ for the cavity can be obtained
  from the finesses $\mathcal{F}_i$ associated to different decay
channels, which can be assumed to be independent such that
\begin{equation}\label{finesse_tot}
\frac{1}{\mathcal{F}}=\sum_i{\frac{1}{\mathcal{F}_i}}\,.
\end{equation}
For each loss channel, the finesse $\mathcal{F}_i$ can be
  determined from the ratio of the power circulating in the cavity
  $P_c$ to the fraction of power that it lost through the respective
  channel $P_i$, using that the time-averaged stored energy and free
  spectral range are given, respectively, by $\langle
  U\rangle = P_c(n_{\mathrm{c}}L_{\mathrm{c}}/c)$ and $\Delta\omega=2\pi
  c/n_{\mathrm{c}}L_{\mathrm{c}}$. Thus, we arrive at
\begin{equation}\label{finesse_i}
  \frac{1}{\mathcal{F}_i}=\frac{P_i}{\Delta\omega\langle U\rangle}
=\frac{P_i}{2\pi P_c}\,.
\end{equation}
The loss channels that will be considered subsequently are (1)
scattering by the ``bulk'' of the electrodes modeled as a single
metallic cylinder , (2) scattering by the ``gap'' between the
electrodes and (3) absorption. The independence between the
  contributions (1) and (2) assumed in Eq.~\eqref{finesse_tot} amounts
  to neglecting the interference between them which is permissible
  when estimating an upper bound.

\paragraph{Incident field and circulating power}

For our calculations, we introduce new cylindrical co-ordinates
$(r',\varphi',z')$ for the electrode with the $z'$-direction
  along its axis. We first express the cavity field that is
incident on the electrode in these primed coordinates. The
result will later be used to determine the scattered and absorbed
fractions of the incident power.

\begin{figure}
  \centering\includegraphics[width=0.4\textwidth]{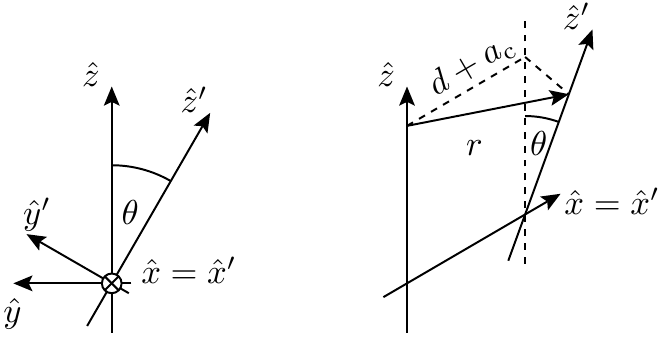}
  \caption{Relative alignment of electrode coordinates
    $(r',\varphi',z')$ and waveguide coordinates
      $(r,\varphi,z)$. The misalignment is determined by a
    small angle $\theta$.}
\label{fig:coordinates_electrode_wg}
\end{figure}

To this end, we consider the projection of the electric field
determined in Section \ref{sec:coupling} onto the electrode's axis, see
  Fig.~\ref{fig:coordinates_electrode_wg},
\begin{equation}
 E_{z'}^{(\mathrm{in})}(z')=\left.\hat z'\cdot\hat\varphi \,\,
   E_{\varphi}\right\vert_{r'=0}\,, 
\label{field_projection}
\end{equation}
which completely determines the losses in the small radius regime
considered.  The origin of the primed axis lies at $(d+\ac,0,0)$ and
the relevant unit vectors are given by
\begin{eqnarray}
\label{coordinate_rel1}
 \hat z' &=& \sin\theta\hat y+\cos\theta\hat z\\
\hat\varphi&=&-\sin\varphi\hat x+\cos\varphi\hat y\,.
\end{eqnarray}
For points on the $z'$-axis we have 
\begin{eqnarray}
 \cos\varphi&=&\frac{d+\ac}{r}\,,\\
r&=&\sqrt{(d+\ac)^2+z'^2\sin^2\theta}\,,\\
z&=&z'\cos\theta\,.
\label{coordinate_rel2}
\end{eqnarray}
For our further calculations it is convenient to express the incident
field via its Fourier transform
$E_{z'}^{(\mathrm{in})}(k')=\int_{-\infty}^{\infty}E_{z'}^{(\mathrm{in})}(z')\ehoch{-\I
  k'z'}{\rm d}z'$. Using this and equations
(\ref{coordinate_rel1})-(\ref{coordinate_rel2}) in equations
\eqref{E_phi} and \eqref{field_projection} we arrive at
\begin{align}
  \nonumber E_{z'}^{(\mathrm{in})}(\tilde{k}) = &\frac{-\I
    \omega_{\mathrm{c}}}{\kappa_{\bot}}s\sqrt{\ac(d+\ac)}
\tilde\xi\ehoch{-\kappa_{\bot}d}
B_{z}(0)\,\times\\
&\times\underbrace{\int\limits_{-\infty}^{\infty}{\rm
    d}x\frac{\ehoch{-\kappa_{\bot}(d+\ac)(\sqrt{1+x^2}-1+\I\tilde
      kx)}}{(1+x^2)^{3/4}}}_{F}\,,
\label{Ez_in_Fourier}
\end{align}
where we have substituted $x=z'\sin\theta/(d+\ac)$ and $\tilde
k=(k'-n_{\mathrm{c}}k\cos\theta)/(\kappa_{\bot}\sin\theta)$. One can
find an approximation for the integral $F$ by applying the method of
steepest descents, using $\kappa_{\bot}(d+\ac)\gg1$ and
$d\sim\kappa_{\bot}^{-1}\ll \ac$, which for $|\theta|\ll1$ yields
\begin{equation}\label{F} 
  |F|\approx\sqrt{\frac{2\pi}{(d+\ac)\kappa_{\bot}}}\ehoch{-\kappa_{\bot}(d+\ac)|
 \tilde k|} 
\end{equation}

Finally, we determine the circulating power to leading order in
  $1/\kappa_\perp\ac$, which is given by
\begin{align}
  P_c&=\int{\rm d}A\cdot \langle \vec S\rangle=\, \frac{\pi
    n_{\mathrm{c}} c}{\text{\textmu}_0}\left(\frac{k}{\gamma}
  \right)^2\left|B_{z}(0)\right|^2\int\limits_0^{\ac}rJ_1^2(\gamma
  r){\rm d}r
  \nonumber\\
  &=\frac{\pi}{2} J_2^2(x_{1,1})
  \frac{cn_{\mathrm{c}}}{\text{\textmu}_0}
  \left(\frac{k\ac}{\gamma}\right)^2\left|B_{z}(0)\right|^2
  \nonumber\\ &\approx0.25 \frac{cn_{\mathrm{c}}}{\text{\textmu}_0}
  \left(\frac{k\ac}{\gamma}\right)^2\left|B_{z}(0)\right|^2,
\label{inpower}
\end{align}
where we have used Eqs.~\eqref{inside} and \eqref{axial}, that
$\pi\int\limits_0^{1}xJ_1^2(x_{1,1}x){\rm d}x =\frac{\pi}{2}
J_2^2(x_{1,1})$, and the time-averaged Poynting vector $\langle
\hat{z}\cdot \vec
S\rangle=\frac{1}{2}\Re\{E_{\varphi}^*B_{r}\}/\text{\textmu}_0$ with
the vacuum permeability \textmu$_0$.

\paragraph{Scattering losses}

Here, we model the electrodes as a single metallic cylinder which for
simplicity is assumed to be perfectly conducting since this maximizes
the scattering and, thus, provides an estimate of an upper bound to
the corresponding losses that is independent of material properties
---naturally for the transparent electrode scenario considered below,
in Section \ref{electrode losses}.d, these losses would be
substantially smaller than this upper bound.  We expand the scattered
field into solutions of the wave equation in cylindrical coordinates
for the electrode. As the radius of the electrode $R'$ is much smaller
than the wavelength of the incident field $\lambda_{\mathrm{c}}$, all
contributions to the scattered power are suppressed at least like $(k
R')^4$, except for s-wave scattering of TM modes, for which the
suppression is only logarithmic. This can be understood in terms of
the Taylor expansions of the corresponding cylindrical harmonics and
the incident field. In turn, to determine the TM s-wave scattering to
leading order in $k R'$, the incident field can be assumed to be
constant for a given cross section and determined by the field at the
electrode's center $E_{z'}^{(\mathrm{in})}(z')$. We neglect multiple
scattering between the waveguide and the electrode and ignore the
dielectric substrate of the latter. Thus, the scattered field
$E_{z'}^{(s)}$ is determined from the homogeneous boundary condition
at the surface of the electrode
$E_{z'}|_{r'=R'}=(E_{z'}^{(\mathrm{in})}+E_{z'}^{(s)}+E_{z'}^{(e)})|_{r'=R'}=0$,
and an outgoing-wave boundary condition at infinity for
  $E_{z'}^{(s)}$ ---here $E_{z'}^{(e)}$ is the evanescent
contribution.

The transverse fields of an outgoing TM solution with
$z'$-dependence $\exp\left(\I k'z'\right)$ are given by
\begin{eqnarray}\label{E_transverse}
\vec E_{\bot}^{(s)}&=&\I \frac{k'}{k^2-k'^2}\nabla_{\bot}E_{z'}^{(s)}\,,\\
\vec H_{\bot}^{(s)}&=&c\epsilon_0 \frac{k}{k'}\hat z'\times\vec E_{\bot}^{(s)}\,,
\end{eqnarray}
with $k'^2<k^2$. For $k'^2>k^2$ the solution is evanescent and does
not contribute to the scattered power. For s-wave scattering
$E_{z'}^{(s)}(k')\propto H_0^{(1)}(k_{\bot}r')\ehoch{\I k'z'}$, where
$k_{\bot}^2=k^2-k'^2$, and to leading order in $k R'$, the scattered
field can be written as
\begin{equation}\label{s-wave}
  E_{z'}^{(s)}(r',\varphi',z')\approx-\int\limits_{-k}^k\frac{{\rm
      d}k'}{2\pi}
  E_{z'}^{(\mathrm{in})}(k')\frac{H_0^{(1)}(\sqrt{k^2\!-\!k'^2}r')}{\I\frac{2}{\pi}
    \ln\left(\sqrt{k^2\!-\!k'^2}R'\right)}\ehoch{ik'z'}\,,
\end{equation}
where $E_{z'}^{(\mathrm{in})}(k')$ is the Fourier transform of the
incident field $E_{z'}^{(\mathrm{in})}(z')$ and we have used the
approximation $H_0^{(1)}(x)\approx\I (2/\pi)\ln x$ for
$|x|\ll1$. We calculate the scattered power by integrating
the energy flux across a cylinder coaxial with the electrode with
  radius $R_*\rightarrow \infty$. Thus, from
  Eqs.~\eqref{E_transverse}-\eqref{s-wave} we obtain for the scattered
  power
\begin{align}
 \nonumber
 P_{\mathrm{s}}\approx&\,\frac{\pi}{4}c\epsilon_0\int\limits_{-k}^{k}{\rm
   d}k'\frac{k\left|E_{z'}^{(\mathrm{in})}(k')\right|^2}{(k^2-k'^2)
\ln^2(\sqrt{k^2-k'^2} R')}\\ 
\nonumber < &\,\pi c\epsilon_0\text{max}\left.\left
\{\left|E_{z'}^{(\mathrm{in})}(k')\right|^2\right\}\right\vert_{|k'|\leq k}\\
&\times\underbrace{2\int\limits_0^k{\rm d}k'\frac{k}{(k^2-k'^2)
\ln^2\left[(k^2-k'^2){R'}^2\right]}}_{G}
\label{scatpower}
\end{align}
where we have used 
\begin{equation}\nonumber
\langle\hat r'\cdot \vec S\rangle=
  \frac{1}{2}\Re\{\hat r'\cdot({\vec E}^*\times\vec
  H)\}=\frac{1}{2}\Re\left\{\frac{\I c\epsilon_0k}{k'^2-k^2}\frac{\partial
    E_{z'}^{(s)}}{\partial r'}{E_{z'}^{(s)}}^*\right\} 
\end{equation}
and $H_0^{(1)}(x)\approx\sqrt{2/\pi x}\exp\left[\I(x-\pi/4)\right]$
for $|x|\gg 1$.  The integral $G$ can be estimated by performing
the substitution $v=\ln[(k-k')/2k]/\ln(2kR)$ and considering
$kR\ll1$, which yields $G\approx 1/(2|\ln(2kR')|)$.  

Hence, from Eqs.~\eqref{finesse_i},
  \eqref{Ez_in_Fourier}-\eqref{inpower}, \eqref{scatpower} and
  \eqref{xi}, and using $\sin\theta\approx\theta$ and
$\cos\theta\approx1$ for $|\theta|\ll1$, and $\kappa_\perp\approx
  \sqrt{n_{\mathrm{c}}^2-1}k$ we arrive at
\begin{equation}
  \frac{1}{\mathcal{F}_s}<
  \frac{\ehoch{-2k\left[d\sqrt{n_{\mathrm{c}}^2-1}+(d+\ac)
        \frac{n_{\mathrm{c}}-1}{|\theta|}\right]}}{n_{\mathrm{c}}\left(k\ac
      \sqrt{n_{\mathrm{c}}^2-1}\right)^3|\ln2kR'|}\,.  
\end{equation}
If we now use that for relevant parameters $d\ll\ac$,
$\ac>\lambda_{\mathrm{c}}$ and
$|\theta|<\sqrt{\frac{n_{\mathrm{c}}-1}{n_{\mathrm{c}}+1}}$, we find
a lower bound for the finesse associated to scattering losses,
\begin{equation}
  \mathcal{F}_{\mathrm{s}}>16\,n_{\mathrm{c}}\left(n_{\mathrm{c}}^2
      -1\right)^{3/2}\ln\left(\frac{\lambda_{\mathrm{c}}}{4\pi R'}\right)
  \ehoch{4\pi\frac{n_{\mathrm{c}}-1}{|\theta|}}\,,
\end{equation}
which is independent of the ratio $\ac/\lambda_{\mathrm{c}}$. Thus,
for $\lambda=\unit{1.1}{\micro\meter}$,
$R'\leq\unit{10}{\nano\meter}$, $n_{\mathrm{c}}=1.44$ and
$\theta\leq\unit{10}{\degree}$, we find
$\mathcal{F}_{\mathrm{s}}>3\times10^{15}$. Hence losses due to
scattering are clearly sufficiently suppressed for electrodes of
subwavelength radius that are approximately aligned with the cavity
rim.

\paragraph{Gap contribution to scattering losses}

So far we have considered scattering from one single continuous
electrode. Actually, our setup comprises two electrodes, see figure
\ref{fig:setup}, separated by a gap in which the nanomechanical
oscillator is positioned. We now estimate an upper bound to the
additional losses that this gap may induce. Here we assume as before
$\theta\ll1$ so that the gap is much smaller than the distance over
which substantial currents are induced in the electrodes and one may
consider $\theta=0$ for this estimate. Denoting the direction joining
the electrodes by $\hat{z}"$, we focus on the relevant regime
$D\gg2R'$ so that one can assume that only the incident field
component along $\hat{z}"$ is screened by them. We model the gap of
size $D$ as a perfectly conducting sphere with radius $D/2$ subject to
an external field along $\hat{z}"$ determined by ${\vec
  E}^{(\mathrm{in})}\cdot\hat{z}"=E_\varphi^{(\mathrm{in})}\cos\theta'
\cos\varphi'$. This should provide an upper bound for the magnitude of
the total induced dipole $p$, which in turn determines the leading
contribution in $kD$ to the scattered power. Thus we find
$p\sim\frac{\pi}{2}\epsilon_0D^3|E_{\varphi}^{(\mathrm{in})}|\cos\theta'_*
\cos\varphi'_*$ \cite{Jackson8} for optimal placement, which yields
for the scattered power
\begin{equation}
  P_{\mathrm{g}}\lesssim\frac{\pi}{48}\epsilon_0ck^4D^6 \left[
    \frac{\omega_{\mathrm{c}}\xi}{\gamma}\left|B_{z}(0)\right|J_1(x_*)\right]^2
  \ehoch{-2\kappa_{\bot}d}\cos^2\theta'_*\,.
\end{equation}
Here we have used Eq.~\eqref{E_phi}, $d'\ll\ac$ and
  $\cos\varphi'_*\approx1$. Along the same lines as before, using
Eqs.~\eqref{inpower} and \eqref{finesse_i} and $\sin^2\theta'_*=2/3$,
  the associated finesse reads
\begin{equation}
  \mathcal{F}_{\mathrm{g}}\gtrsim 0.8\,\frac{n_{\mathrm{c}}\ac^4
    \lambda_{\mathrm{c}}^2}{D^6\xi^2}
  \left(n_{\mathrm{c}}^2-1\right)\ehoch{\frac{4\pi
      d}{\lambda_{\mathrm{c}}}
    \sqrt{n_{\mathrm{c}}^2-1}}
\end{equation}
which yields $\mathcal{F}_g\gtrsim 3 \times 10^{9}$ for the parameters
discussed in Section \ref{sec:setup} and $D=40\,$nm. Thus, additional
scattering losses due to the gap between the electrodes are also
negligible.

\paragraph{Absorption losses}

We assume here transparent electrodes afforded by cylindrical shells
with 2D conductivity $\sigma$ that absorb the power
\begin{equation}
P_{\mathrm{a}}=\frac{1}{2}\int\limits_{-\infty}^{\infty}{\rm d}z'
\Re\left\{I^*E_{z'}^{(\mathrm{in})}\right\}\,,
\label{absorbed_power}
\end{equation}
since $E_{z'}^{(\mathrm{in})}$ induces a current $I=2\pi
R'\sigma E_{z'}^{(\mathrm{in})}$ in each electrode
---here, as in \ref{electrode losses}.b., we neglect the small gap. From
  Eqs.~\eqref{E_phi} and
  \eqref{field_projection}-\eqref{coordinate_rel2} we obtain for the
  absorbed power
\begin{align}\label{absorbed}
  P_{\mathrm{a}}\approx&\,\pi
  R'\sigma\xi^2\ac\sin\theta\left[\frac{\omega_{\mathrm{c}}}{\gamma}
J_1(x_*)\right]^2 \ehoch{-2\kappa_{\bot}d}|B_{z}(0)|^2\,\times\\
    &\times\underbrace{\int
\limits_{-\infty}^{\infty}{\rm d}x\frac{\ehoch{-2\kappa_{\bot}(d+\ac)
(\sqrt{1+x^2}-1)}}{(1+x^2)^{3/2}}}_{J}\nn
\end{align}
where we have again substituted $x=z'\sin\theta/(d+\ac)$. Using
  the method of steepest descents we estimate
$J\approx(\kappa_{\bot}(d+\ac)/\pi)^{-1/2}$ for
$2\kappa_{\bot}(d+\ac)\gg1$. Thus using $d\ll \ac$ and
  Eqs.~\eqref{finesse_i}, \eqref{inpower} and \eqref{absorbed},
yields for the finesse associated to absorption
\begin{equation}
\frac{1}{\mathcal{F}_{\mathrm{a}}}\approx\frac{\sigma
  R'\xi^2}{\pi n_{\mathrm{c}}c\epsilon_0\ac}\sqrt{\frac{\pi}{\kappa_{\bot}\ac}}
\ehoch{-2\kappa_{\bot}d}\sin\theta \left[\frac{J_1(x_*)}{J_2(x_{1,1})}\right]^2\,.
\end{equation} 
We consider now the specific case where the transparent electrodes are
provided by a pair of nanotubes. The latter exhibit a maximum in
the conductivity $\sigma_{\mathrm{max}}=8\mathrm{e}^2/\mathrm{h}$
\cite{nnano.2010.248}. By assuming an off-resonant
$\sigma=\tilde\sigma\sigma_{\mathrm{max}}$ with $\tilde\sigma < 1$, we
get
\begin{equation}
  \mathcal{F}_{\mathrm{a}}\approx\frac{\sqrt{\pi}n_{\mathrm{c}}\ac}{
    16\alpha_{\mathrm{F}}\tilde\sigma R'\sin\theta}
\left(\kappa_{\bot}\ac\right)^{5/2}\ehoch{2\kappa_{\bot}d}\,,
\end{equation}
with the fine structure constant $\alpha_{\mathrm{F}}\approx1/137$. If
we consider the same values as before except that now
$\theta<\unit{3}{\degree}$ and $R'=\unit{2.5}{\nano\meter}$ and
assuming $\tilde\sigma<1/20$ we find $\mathcal{F}_{\mathrm{a}}\gtrsim
3 \times 10^{8}$.  Hence even though absorption losses clearly
dominate over scattering losses, their effect can still be neglected
for electrode radii $R'<\unit{3}{\nano\meter}$ and alignment angles
$\theta<\unit{3}{\degree}$.

\subsection{Electrical noise}
\label{Electrical noise}

Here we give estimates of decoherence rates for the
nanoresonator induced by noise in the inhomogeneous
  electric fields. Such noise might originate from voltage
fluctuations $\delta U$ due to the electrodes resistance
(Johnson-Nyquist noise) or from moving charges on the chip surface
($1/f$-noise). We calculate the respective single-phonon
decoherence rates $\Gamma_{\delta U}$ and $\Gamma_{1/f}$ from the
corresponding noise spectra $S_{\delta F_{\delta U}}$ and $S_{\delta
  F_{1/f}}$ using the relation
\begin{equation}
 \Gamma_i\sim\frac{\xzpm^2}{\hbar^2}S_{\delta F_i}(\omega_{\mathrm{m}})
\end{equation}
with
\begin{equation}
S_{\delta F_i}(\omega)=\text{Re}\int\limits_0^{\infty}{\rm
  d}\tau\left\langle 
\delta F_i(\tau)\delta F_i(0)+\delta F_i(0)\delta
F_i(\tau)\right\rangle\ehoch{\I\omega\tau}\,,
\end{equation}
where $\delta F_i$ is the force fluctuation acting on the
resonator. The electrostatic gradient force acting on a resonator can
be expressed by
\begin{equation}
 F_{\mathrm{el.}}= \frac{\alpha}{2} \, \frac{\partial}{\partial X} 
\int E^{2} {\rm d}l \sim \alpha a E\left(\frac{E}{a}\right)\,,
\end{equation}
where we have estimated the
field gradient at a distance $a$ from the electrode by $E/a$ and used
the fact that the field mainly acts on the nanotube in a region of
length $a\ll L$. For a field with fluctuations associated to different
independent sources $E+\sum_i\delta E_i$, the force fluctuations are
then given by
\begin{equation}
\delta F_i \sim 2\alpha E \delta E_i\,.
\end{equation}
Thus, the resulting decoherence rates read
\begin{equation}
  \Gamma_i\sim\frac{\xzpm^2}{\hbar^2}S_{\delta
    F_i}\sim4\frac{\xzpm^2}{\hbar^2}
  \alpha^2E^2S_{\delta E_i}\,,
\end{equation}
where the $S_{\delta E_i}$ are the noise spectra for the different
electric field fluctuations.

\paragraph{Johnson-Nyquist noise}
For Johnson-Nyquist noise \cite{Nyquist}, we have fluctuating voltages
$\delta U$ with
\begin{equation}
S_{\delta U}=4k_{\mathrm{B}}T R_e \quad \text{and thus} 
\quad  S_{\delta E}\sim S_{\delta U}/a^2\,,
\end{equation}
for an ambient temperature $T$ and an internal resistance
$R_e$. For our setup we find $\Gamma_{\delta U}/R_e \lesssim
\unit{10^{-2}}{\hertz\per\ohm}$ at $T=\unit{20}{\milli\kelvin}$, which
is well below the relevant mechanical decoherence rate $\gammam
\overline{n}\approx\unit{0.1}{\kilo\hertz}$ for relevant
  resistances $R_e\lesssim\unit{1}{\ohm}$.

\paragraph{$1/f$-noise}
The origin of $1/f$-noise is usually associated with surface charge
fluctuations in the device. An electric field noise density
$S_{E}(\omega/2\pi=\unit{3.9}{\kilo\hertz})\approx\unit{4}{\square\volt
  \rpsquare\meter\reciprocal\hertz}$ has been measured at
$T=\unit{300}{\kelvin}$ and at a distance of $\unit{20}{\nano\meter}$
between a charged resonator and a gold surface \cite{Stipe}. For a
scaling $S_{E}(\omega)\sim T/\omega$ \cite{Stipe,Rabl} this
corresponds to
$S_{E}\approx\unit{2\times10^{-7}}{\square\volt\rpsquare\meter\reciprocal\hertz}$
for our conditions with $T=\unit{20}{\milli\kelvin}$ and
$\omega_{\mathrm{m}}/2\pi\approx\unit{5.2}{\mega\hertz}$. Thus we
  expect for the associated decoherence rate
$\Gamma_{1/f}\lesssim\unit{0.15}{\hertz}$, which is again well
below the mechanical decoherence rate $\gammam \overline{n}$.

These results are also corroborated by recent estimates that were
obtained for a related setup \cite{Wilson-Rae09}.

%\section{Effective dynamics for the nanobeam(s)}
\section{Control mechanisms and applications}
\label{sec:controlmech}

The Hamiltonian \eqref{hamiltonian_shifted} of the full optomechanical
system with tuned nanobeams potentially leads to complex dynamics for
photons and phonons. Here, we focus on scenarios where driven cavity
modes and suitable electric gradient fields are used to control the
dynamics of one or several nanobeams. We summarize the basic
principles for three conceptually different schemes, namely (i) the
selective addressing of transitions in the mechanical spectrum by
cavity sideband driving, (ii) the coherent interaction between several
nonlinear nanobeams mediated by a common driven cavity mode and (iii) the
manipulation of a single resonator's state by time dependent gradient
fields.

\begin{figure}
\centering
\hspace{-0.9cm}
 \includegraphics[height=0.45\columnwidth]{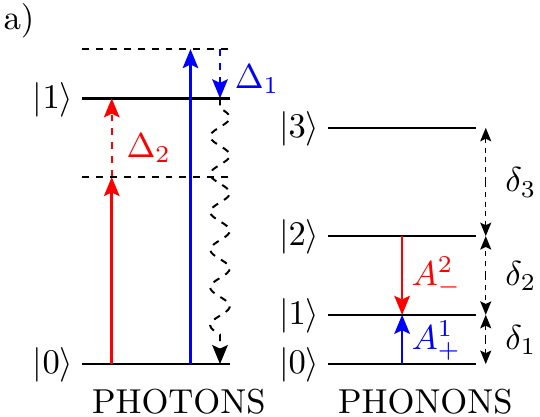} \\
\vspace{0.6cm}
 \includegraphics[height=0.45\columnwidth]{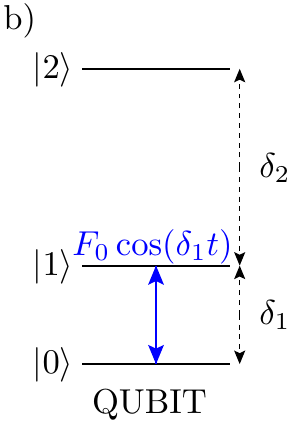}
 \includegraphics[height=0.45\columnwidth]{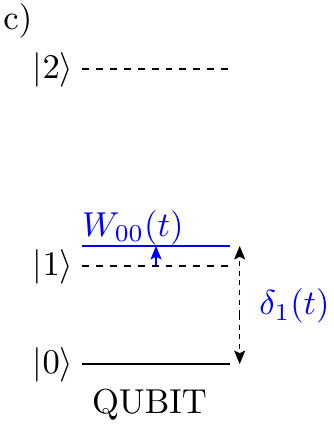}
 \caption{Different schemes for resonant interactions. a) Interaction
   between mechanical excitations and detuned photons entering the
   cavity. b) and c) interaction between a classical gradient field
   and mechanical excitations.}
 \label{fig:Schemes}
\end{figure}

The first scheme represents a suitable extension of the standard
sideband cooling technique
\cite{PhysRevLett.99.093901,PhysRevLett.99.093902} to nonlinear
resonators. Here, the detuning of a red (blue) detuned laser drive is
only resonant with one specific transition
$\ket{n}\rightarrow\ket{n-1}$ ($\ket{n}\rightarrow\ket{n+1}$) in the
nonlinear mechanical spectrum, see figure
\ref{fig:Schemes}a. Therefore, if the mechanical nonlinearity is
resolved by the cavity linewidth, $\lambda>\kappa$, appropriate laser
drives can lead to highly nonclassical steady states for the
mechanical motion. For a single nanobeam for example, this allows for
the preparation of stationary Fock states with high fidelity
\cite{Rips2012}.  The sideband driving technique could potentially
also be applied to more complicated level structures, for example the
collective modes of several interacting nanobeams, and thus
constitutes a versatile control mechanism.

The second scheme uses the cavity to mediate a coherent coupling
between several nanobeams that all couple to the same photon
mode. Here, the photon mode is driven with a large detuning to be
off-resonant to any mechanical transition frequency. The coherent
photon background field that builds up inside the cavity leads to an
effective interaction $\sim\X_i\X_j$ between any pairs $i,j$ of
nanobeams. In order to exchange excitations via this coupling, proper
resonance conditions have to be met. By tuning each of the nanobeams
using their respective electrodes, interactions between desired pairs
of beams can be realized \cite{Rips2013}. Furthermore, due to
the nonlinear spectra, it is possible to restrict the dynamics of the
nanobeams to the ``qubit'' subspace built up by the states
$\left\{\ket{0},\ket{1}\right\}$, c.f. equation \eqref{eq:diagHm}, for
each resonator.

Finally, beside the static tuning capability, see equation \eqref{Fn},
the gradient fields provided by the tip electrodes can be used to
perform coherent operations on any nanobeam. This becomes most obvious
if one considers the qubit subspace $\{\ket{0},\ket{1}\}$ for one
nanobeam. Here, a drive $F_0(t)\propto\cos(\delta_1 t)$, where
$\delta_1=(E_1-E_0)/\hbar$ is the qubit transition frequency,
implements a $\sigma_x$ rotation, see figure \ref{fig:Schemes}b. A
temporary shift of the qubit transition frequency $\delta_1$ can be
achieved by a temporary $W_{00}(t)$ contribution, which corresponds to
a $\sigma_{z}$ rotation, see figure \ref{fig:Schemes}c. Note that the
drive $F_0(t)$ associated with the tip electrodes is a coherent drive,
while the cavity sideband driving technique constitutes a stochastic
drive. Together with the coherent coupling of several nanobeams, the
time dependent gradient fields can for example be employed to build up
a universal set of quantum gates for quantum information processing
\cite{Rips2013}.

\section{Measurement via output power spectrum}
\label{sec:measurement}

\begin{figure*}
\centering
 \includegraphics[width=0.9\textwidth]{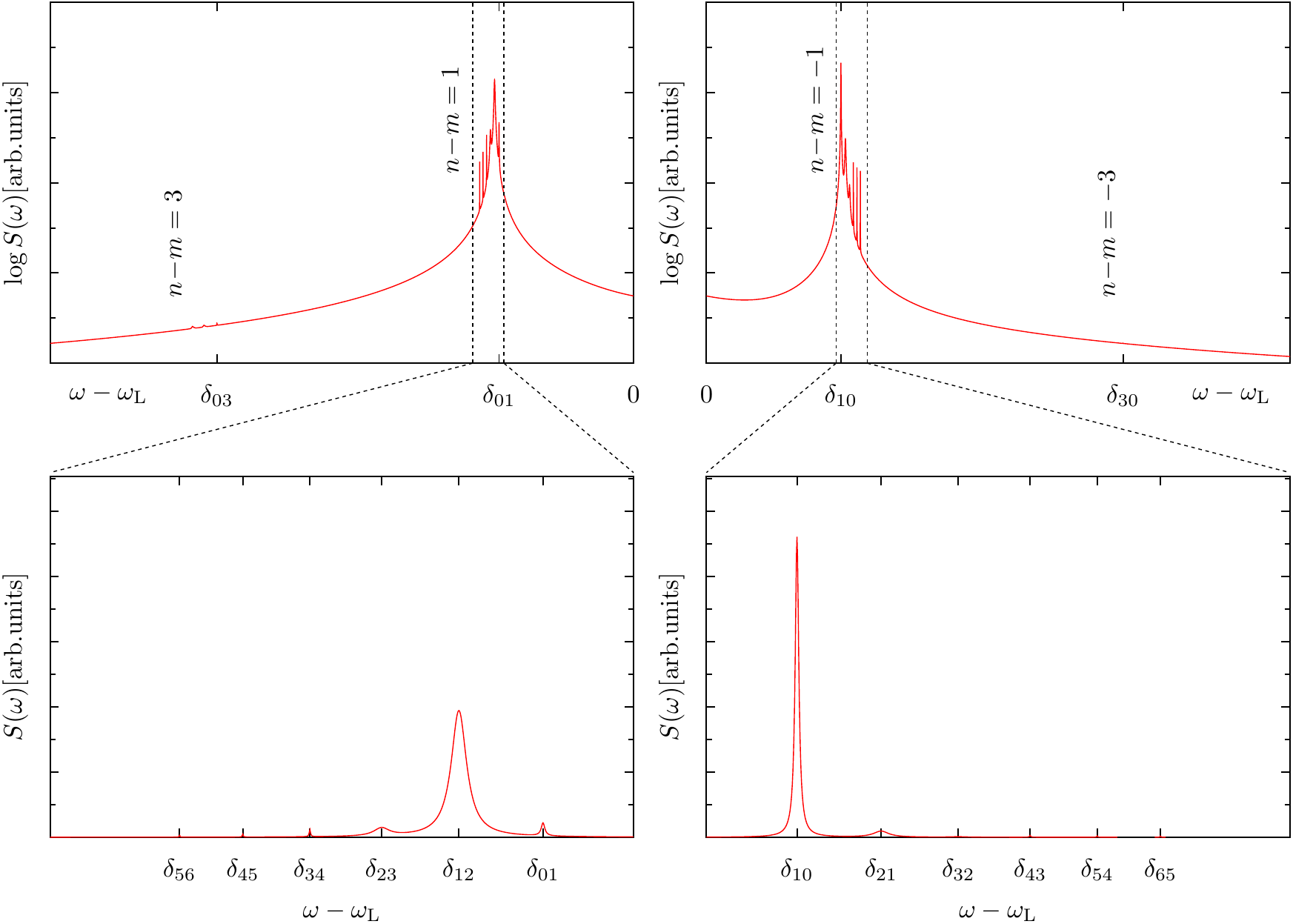}
 \caption{Output spectrum characterizing a motional state of a single
   nanoresonator where only the lowest few states are populated.
   Here, $P_{0} = 0.0391$, $P_{1} = 0.9137$, $P_{2} = 0.0430$, $P_{3} = 0.0024$,
   $P_{4} = 0.0006$, $P_{5} = 0.0003$ and $P_{j} <
     0.0003$ for $j>5$. Due to the parity of $\X$,
   there are only sideband groups at frequencies that correspond to
   processes where an odd number of phonons is scattered. The
   sidebands drop quickly in intensity with growing distance to the
   main line. The fine structure of the $n\!-\!m=\pm1$ groups gives
   information about the level population of the resonator. The
   frequency axis is labeled in terms of the mechanical transition
   frequencies $\delta_{nm}$, and a peak at
   $\omega-\omega_{\mathrm{L}}=\delta_{mn}$ with $\delta_{mn}>0$
   ($\delta_{mn}<0$) corresponds to anti-Stokes (Stokes) photon
   scattering events that occur at a rate proportional to the
   occupation of the respective initial mechanical level. In this
   example the level spacings correspond to a $\unit{1}{\micro\meter}$
   length $(10,0)$ CNT that is softened so that
   $\omega_{\mathrm{m}}/2\pi=\unit{5.23}{\mega\hertz}$, which leads to
   $\delta_{10}\approx\lambda'=\unit{209}{\kilo\hertz}$. Three
     ``active'' state-preparation lasers with detunings
     $\Delta_1\approx\delta_{10}$,
     $\Delta_{2}\approx\delta_{12}$ and 
     $\Delta_{3}\approx\delta_{23}$ induce optomechanical
     couplings
     $g_{\mathrm{m},1}=g_{\mathrm{m},2}=g_{\mathrm{m},3}=\unit{20.9}{\kilo\hertz}$
   ---for the parameters considered (cf.~Section
   \ref{sec:setup}) this value can be achieved keeping the intracavity
   absorption in the microwatt range \cite{Rips2012}. The resulting
   optomechanical interactions with the nanobeam dominate over the
   intrinsic mechanical decoherence rates and determine the broadening
   of the levels involved, see section \ref{appendix B} ---the
   remaining relevant parameters are:
   $\kappa_\mathrm{ex}/\kappa=0.1$,
     $\kappa_1/2\pi=\kappa_2/2\pi=\kappa_3/2\pi=52.3\,$kHz,
     $\omega_{\mathrm{m}}/\gamma=5\times10^6$ and $T=20\,$mK.}
 \label{fig:Spectrum}
\end{figure*}

The steady state of a nanoresonator can be probed with an additional
laser, weakly driving one cavity mode on resonance, i.e. with
$\Delta=0$. Then, information about the state of the mechanical
resonator can be extracted from the sideband structures in the power
spectrum, which correspond to photons that have been up- or down
converted during the interaction with the mechanical motion. The
intensity of the sideband peaks depends on the population of the
mechanical energy eigenstates. Thus, the power spectrum only provides
information about the diagonal entries $P_{n}$ of the density matrix
describing the mechanical resonator represented in the basis formed by
eigenstates of the Hamiltonian $H_{\mathrm{m}}$, see
Eq.~(\ref{eq:diagHm}). For a probe laser resonantly driving a cavity
mode at frequency $\omega_{\mathrm{L}}$, the spectrum shows a sideband
structure
\begin{equation}
S(\omega)=\sum\limits_{nm}\frac{\kappa_{\mathrm{ex}}|g_{\mathrm{m}}|^2}{4\delta_{nm}^2
+\kappa^2}X_{nm}^2\text{L}_{nm}(\omega)P_n\,,
\label{spectrum2}
\end{equation}
with Lorentzian sideband peaks determined by 
\begin{equation}
  \text{L}_{nm}(\omega)=\frac{1}{\pi}\frac{\gammaeff^{nm}/2}{\left[\omega-
      \omega_{\mathrm{L}}-\delta_{nm}\right]^2+(\gammaeff^{nm}/2)^2}\,,
\end{equation}
with $\gammaeff$ as given in equation (\ref{gammaeff}), see appendix
\ref{appendix B} for details. Here, $g_{\mathrm{m}}$ is the
optomechanical coupling strength associated with the probe laser and
the $\delta_{nm}=(E_n-E_m)/\hbar$ denote the mechanical transition
frequencies. The peaks appear in groups with $n\!-\!m=1,3,...$;
see figure \ref{fig:Spectrum}. The occupation probabilities $P_n$
for the eigenstates $\ket{n}$ can be extracted from the peak
intensities within the main sidebands with $n\!-\!m=1$
\cite{Rips2012}.

For the readout of mechanical qubits as discussed in \cite{Rips2013},
a shelving technique can be used to determine whether a qubit is in
state $\ket{0}$ or in state $\ket{1}$. Here, a balanced cycling
transition between $\ket{1}\leftrightarrow\ket{2}$ using a cooling
laser on $\ket{2}\rightarrow\ket{1}$ and a coherent rf-drive with
local gradient fields on those two levels causes a continuous stream
of up converted photons only if the resonator is found in the state
$\ket{1}$. This can be detected by measuring the corresponding
sideband spectrum. Here, a large enough number of photons has to be
collected before external damping destroys the intermediate state,
which requires
$\kappa_{\mathrm{ex}}|g_{\mathrm{m}}|^2/\kappa^2\gg\gamma\overline n$.

\section{Conclusions and Outlook}
\label{sec:outlook}
We have introduced a scheme to access a novel regime of optomechanics
where the motion of the nanomechanical resonator becomes anharmonic
and thus allows to explore genuine quantum dynamics.  In our approach
inhomogeneous electrostatic fields are applied to the nanomechanical
resonator to enhance its anharmonicity per phonon until it becomes
comparable to the linewidth of a high finesse optical cavity.  For
realistic experimental conditions, sufficiently large optomechanical
couplings can be realized and losses induced by the tip electrodes can
be suppressed to a negligible level.  Furthermore populations of the
energy eigenstates of such nonlinear mechanical oscillators can be
extracted from the output spectrum of a probe laser.  The approach
thus paves the way towards exploring nonclassical dynamics of
nanomechanical oscillators at the single-phonon level.

\section{Acknowledgements}
This work is part of the Emmy Noether project HA 5593/1-1 and was
supported the CRC 631, both funded by the German Science Foundation
(DFG).  IWR acknowledges support by the German Science Foundation
(DFG) via the grant WI 3859/1-1 and Nanosystems Initiative Munich
(NIM).  The authors thank J.~Kotthaus, T.J.~Kippenberg, and
A.~Schliesser for enlightening discussions.

\appendix

\section{Corrections due to nonlinear mode coupling}
\label{appendix A}

In order to estimate the strength of the mode coupling, we rewrite the
nonlinearity \eqref{nonlinearity1} as
\begin{equation}
  \lambda_{ijkl}^0=\frac{\hbar}{32\tilde{\kappa}^2m}\left[\frac{\mu^2L^2
 \tilde M_{ij}\tilde M_{lk}}{\sqrt{m^*_im^*_jm^*_km^*_l}\nu_i\nu_j\nu_k\nu_l}
  \right]\,,
\label{bracket}
\end{equation}
where $m = \mu L$ is the physical mass of the rod.  The term in
  brackets solely depends on the mode shape for the doubly clamped
boundary conditions and is independent of the parameters
  $\tilde{\kappa}$, $\mu$, $L$, of the oscillator. This can be seen
from substituting $\tilde\phi_n(\tilde x)\equiv \phi_n(L\tilde x)$,
which yields
\begin{align}
 \tilde{M}_{ij}&=L M_{ij}=\int_0^1\tilde\phi_i'\tilde\phi_j'{\rm d}\tilde x\,,\\
m^*_n&= \mu L \int_0^1\tilde\phi_n^2{\rm d}\tilde x\,.
\end{align}
Table \ref{couple_vals} shows some numerically obtained values for the
bracket in equation \eqref{bracket} that are relevant for the
fundamental mode.  In the case of an electrostatically tuned
  resonator, the modified nonlinear couplings read
\begin{equation}
 \lambda_{ijkl}=\lambda^0_{ijkl}\sqrt{\zeta_i\zeta_j\zeta_k\zeta_l}\,,
\end{equation}
where $\zeta_i=\omega_{i,0}/\omega_i$ is the factor by which the
frequency of mode $i$ is reduced due to the presence of the gradient
fields. While this factor is usually intended to be larger than unity
for the fundamental mode $\zeta_1\sim10$, it remains close to unity
for the higher modes.

Phonon transfer between modes is strongly suppressed because of
resonance mismatches, as $\lambda_{ijkl}\ll\sum_{n=ijkl}(\pm
\omega_n)$ for processes where the phonon number in each mode is not
preserved. One should note that the relevant frequency ratios scale as
$\lambda_0/\omega_{\mathrm{m},0}\ll1$ and the dominant processes of
this type affecting the fundamental mode involve its coupling to the
next higher mode with the same symmetry, which is the third mode. In
addition, the fundamental mode experiences a modification of its
rigidity due to the thermal and quantum fluctuations of higher order
modes with $\omega_n<\omega_N\sim\csound\pi/L$. This effect however
can be taken into account by a proper redefinition of the fundamental
mode's rigidity.

\begin{table}
%\begin{tabular}{ccc}
 \begin{tabular}{|c||c|c|c|c|c|}
\hline
$j\diagdown i$&$1$&$2$&$3$&$4$&$5$\\
\hline\hline
$1$&0.3024&---&0.1029&---&-0.0512\\
\hline
$2$&---&0.4106&---&-0.0848&---\\
\hline
$3$&0.1029&---&0.4498&---&0.0705\\
\hline
$4$&---&-0.0848&---&0.4721&---\\
\hline
$5$&-0.0512&---&0.0705&---&0.486232\\
\hline
%\end{tabular}&&
\end{tabular}
\caption{$32\tilde{\kappa}^2m\lambda_{11ij}^0/\hbar$. Only pairs
    of modes $i,j$ with the same parity yield a finite coupling
    of this type.} 
\label{couple_vals}
\end{table}

\section{Output power spectrum}
\label{appendix B}

A steady state of the mechanical resonator can be probed via a
resonant laser drive on an additional cavity mode. The quantum motion
of the nanoresonator is described by the reduced master equation
  \cite{Rips2012}, 
\begin{equation}
 \dot\mu\approx-\I\left[\sum\limits_nE_n\ket{n}\bra{n},\mu\right]+\frac{1}{2}\sum\limits_{nm,j}A^{nm}_j\D(\ket{n}\bra{m})\mu+\D_{\gamma}\mu\,.
 \label{eq:reduced_mg}
\end{equation}
Here, $\D_{\gamma}\mu$ includes the external mechanical damping via
standard Lindblad terms and the influence of the lasers is given by
the rates
\begin{equation}
  A^{nm}_j=|\gmj|^2\frac{X_{nm}^2\kappa_j}{4\left(\Delta_j - \delta_{nm}
    \right)^2+\kappa_j^2}\,. 
\end{equation}
Here, for example, $j=0$ labels the probe laser with $\Delta_0=0$ and
the other lasers with $j=1,2,...$ are used for the steady state
preparation, see Section \ref{sec:controlmech}. The rates
$A^{nm}_0$ have to be small, i.e. $A^{nm}_0\lesssim\gamma\overline n$,
to assure a weak measurement. The output power spectrum is given by
\begin{equation}
  S(\omega)\!=\!\sum\limits_j\!\frac{1}{2\pi}\!\!\int\limits_{-\infty}^{\infty}\!\!
  \mathrm{d}\tau\ehoch{-\I(\omega-\omega_{\mathrm{L},j})\tau}\! 
  \left.\expval{a_{\mathrm{out},j}^{\dag}(t+\tau)a_{\mathrm{out},j}(t)}\right|_{\mathrm{SS}}
  \label{power_spectrum}
\end{equation}
where the output fields $a_{\mathrm{out},j}(t)$ are related to the
intracavity fields $a_j(t)$ via the standard input-output relation
\cite{PhysRevA.31.3761},
\begin{equation}\label{inout}
  a_{\mathrm{out},j}=\sqrt{\kappa_{\mathrm{ex}}}a_j+a_{\mathrm{in},j}\,.                      
\end{equation}
Here, we only focus on the output for the probe field and thus
drop the index $j$. The dynamics of the intra cavity field can
be described by a quantum Langevin equation
\begin{align}\nonumber
\dot a =&-\frac{\kappa}{2} a-\I G_0\xzpm\left(\bplus+b\right)\left(a+\alpha\right)\\
&+\sqrt{\kappa_{\mathrm{ex}}}\delta a_{\mathrm{in}}(t)+
\sqrt{\kappa-\kappa_{\mathrm{ex}}}\delta c_{\mathrm{in}}(t)\,,\label{QLE}
\end{align}
where $\delta a_{\mathrm{in}}$ and $\delta c_{\mathrm{in}}$ are the
fluctuations of the input field in the laser mode and the other bath
modes. Defining $a^{(0)}$ to be a solution for $G_0=0$, we can
integrate \eqref{QLE} formally and apply a Dyson series type expansion
to first order in the optomechanical coupling strength $g_{\mathrm{m}}$, to
find for the motion of the cavity modes,
cf. \cite{1367-2630-10-9-095007},
\begin{align}
 \nonumber a(t)&\approx
 a^{(0)}(t)-\I\frac{g_{\mathrm{m}}}{2}\int\limits_0^t\mathrm{d}\tau
\ehoch{-\frac{\kappa}{2}(t-\tau)}\left[\bplus(\tau)+b(\tau)\right]\\
 &\approx
 a^{(0)}(t)-\I\frac{g_{\mathrm{m}}}{2}\sum\limits_{n<m}X_{nm}\left[\frac{\ket{n}
\bra{m}(t)}{\I\delta_{nm}-\kappa/2}+\text{H.c.}\right]\,.\label{dyson}
\end{align}
The contribution of the fluctuations is included in the free field
solution $a^{(0)}(t)$ and the input field operators have already been
written in a shifted representation
$a_{\mathrm{in}}(t)\rightarrow\delta
a_{\mathrm{in}}(t)+\expval{a_{\mathrm{in}}}$, which splits off the
coherent part of the input.  We substitute \eqref{inout} and
\eqref{dyson} into \eqref{power_spectrum} and concentrate on the
contributions to the sidebands, which to lowest order in
$g_{\mathrm{m}}/\om$ are given by the first order two-time
correlations of the mechanical motion. The latter can be calculated
from the reduced master equation \eqref{eq:reduced_mg} using the
quantum regression theorem. We find that
\begin{equation}
  \big{\langle}\ket{n}\bra{m}(t+\tau)\ket{m}\bra{n}(t)\big{\rangle}=
  \ehoch{\left(\I
      \delta_{nm}-\gammaeff^{nm}/2\right)\tau}P_n\,,
\end{equation}
where $P_n=\bra{n}\mu|_{\mathrm{SS}}\ket{n}$ are the
  probabilities to find the resonator in the eigenstate $\ket{n}$,
  are the only nonvanishing contributions. Thus the sideband spectrum
around the probe laser frequency reads
\begin{align}
 S(\omega)=\frac{\kappaex}{2\pi\kappa}\sum_{nm}
\frac{A^{nm}_0\gammaeff^{nm}}{\left(\omega-\omega_{\mathrm{L}}-
\delta_{nm}\right)^2+\left(\gammaeff^{nm}\right)^2/4}P_n\,. 
\end{align}
The resulting peak linewidths
\begin{align}\label{gammaeff}
\nonumber
\gammaeff^{nm}=&\sum\limits_{l,j}\left(A^{lm}_j+A^{ln}_j\right)+\gamma\overline
n\left(\sum\limits_{k>m}X_{km}^2+\sum\limits_{k>n}X_{kn}^2\right)\\ 
&+\gamma(\overline
n+1)\left(\sum\limits_{l<m}X_{lm}^2+\sum\limits_{l<n}X_{ln}^2\right) 
\end{align}
satisfy $\gammaeff^{nm}\ll\lambda$, where the nonlinearity $\lambda$
is the typical peak distance within the fine structure of one
sideband, see figure \ref{fig:Spectrum}.

\end{document}